\documentclass[superscriptaddress]{revtex4-2}

\usepackage{graphicx,graphics}
\usepackage{dcolumn}
\usepackage{amsmath,amssymb,amsfonts}
\usepackage{latexsym,verbatim}
\usepackage{bm}
\usepackage{mathtools}
\usepackage{dsfont}
\usepackage{bbold}
\usepackage{color}
\usepackage{ulem}
\usepackage[breaklinks=false,colorlinks,citecolor=blue,linkcolor=blue,urlcolor=blue]{hyperref}
\graphicspath{{./immagini/}}

\usepackage{subcaption}

\usepackage{comment}

\newcommand{\NB}[1]{{\color{blue}#1}}

\usepackage{braket}
\usepackage{soul}
\usepackage{amsthm}

\DeclarePairedDelimiter\ceil{\lceil}{\rceil}

\graphicspath{{pictures/}}
\usepackage{ulem}
\usepackage{amsmath}
\newtheorem{proposition}{Proposition}
\usepackage{braket}
\usepackage{stmaryrd}
\usepackage{comment}

\begin{document}
\title{Parallelizable Synthesis of Arbitrary Single-Qubit Gates \\with Linear Optics and Time-Frequency Encoding}


\author{Antoine Henry}
\affiliation{ T\'el\'ecom Paris-LTCI, Institut Polytechnique de Paris, 19 Place Marguerite Perey, 91120 Palaiseau, France}
\affiliation{ Centre for Nanosciences and Nanotechnology, CNRS, Universit\'e Paris-Saclay, UMR 9001,10 Boulevard Thomas Gobert, 91120 Palaiseau, France}

\author{Ravi Raghunathan}
\affiliation{ T\'el\'ecom Paris-LTCI, Institut Polytechnique de Paris, 19 Place Marguerite Perey, 91120 Palaiseau, France}

\author{Guillaume Ricard}
\affiliation{ T\'el\'ecom Paris-LTCI, Institut Polytechnique de Paris, 19 Place Marguerite Perey, 91120 Palaiseau, France}

\author{Baptiste Lefaucher}
\affiliation{ T\'el\'ecom Paris-LTCI, Institut Polytechnique de Paris, 19 Place Marguerite Perey, 91120 Palaiseau, France}

\author{Filippo Miatto}
\affiliation{ T\'el\'ecom Paris-LTCI, Institut Polytechnique de Paris, 19 Place Marguerite Perey, 91120 Palaiseau, France}

\author{Nadia Belabas}
\affiliation{ Centre for Nanosciences and Nanotechnology, CNRS, Universit\'e Paris-Saclay, UMR 9001,10 Boulevard Thomas Gobert, 91120 Palaiseau, France}

\author{Isabelle Zaquine}
\affiliation{ T\'el\'ecom Paris-LTCI, Institut Polytechnique de Paris, 19 Place Marguerite Perey, 91120 Palaiseau, France}

\author{Romain All\'eaume}
\affiliation{ T\'el\'ecom Paris-LTCI, Institut Polytechnique de Paris, 19 Place Marguerite Perey, 91120 Palaiseau, France}
\email{romain.alleaume@telecom-paris.fr}

\begin{abstract}
We propose novel methods for the exact synthesis of single-qubit unitaries with high success probability and gate fidelity, considering both time-bin and frequency-bin encodings.
The proposed schemes are experimentally implementable with a spectral linear-optical quantum computation (S-LOQC) platform, composed of electro-optic phase modulators and phase-only programmable filters (pulse shapers). 

We assess the performances in terms of fidelity and probability of the two simplest 3-components configurations for arbitrary gate generation in both encodings and give an exact analytical solution for the synthesis of an arbitrary single-qubit unitary in the time-bin encoding, using a single-tone Radio Frequency (RF) driving of the EOMs.
We further investigate the parallelization of arbitrary single-qubit gates over multiple qubits with a compact experimental setup, both for spectral and temporal encodings.
We systematically evaluate and discuss the impact of the RF bandwidth - that conditions the number of tones driving the modulators - and of the choice of encoding for different targeted gates.
We moreover quantify the number of high fidelity Hadamard gates that can be synthesized in parallel, with minimal and increasing resources in terms of driving RF tones in a realistic system. 
Our analysis positions spectral S-LOQC as a promising platform to conduct massively parallel single qubit operations, with potential applications to quantum metrology and quantum tomography.

    \end{abstract}

    \maketitle
    
    \section{Introduction}
Quantum information processing with light is a promising direction for near-term quantum computing.
 Quantum photonics systems are currently placed at the forefront of the technological race to engineer well-controlled optical quantum states in a  Hilbert space of very high dimensionality, with a record value of $10^{43}$\cite{zhong2021phase}. 
 Another key advantage of  quantum photonics hardware is {scalability} based on photonic integration,  now enabling compact photonic circuits approaching 1,000 components for millimetre-scale footprints \cite{wang2020integrated}.
 
 A drawback of using light is however that photons do not interact with one another, which means that conditional operations need to be mediated by matter via some non-linear process \cite{boyd2020nonlinear}. Even though non-linear processes can theoretically implement the desired operations, this can lead in practice to very inefficient schemes, i.e. with very low probability of success. 
Linear Optics Quantum Computation (LOQC), introduced in the pioneering work of Knill, Laflamme and Milburn in 2001 \cite{knill2001scheme}, performs universal quantum computing with light, using only linear optics and postselection.
 A different scheme, spectral linear-optical quantum
computation (S-LOQC), was proposed \cite{lukens2017frequency} as a highly promising approach for scalable quantum information processing. S-LOQC harnesses photonic qubits encoded over spectral modes (dual-rail frequency encoding) and gate transformations implemented with off-the-shelf telecom components: Electro-Optics Modulators (EOM) and phase-only programmable filters (PS). S-LOQC notably leverages the spectral degrees of freedom that can allow to reach high dimensionality much more easily than spatial and polarization degrees of freedom. It can be used as a resource for parallelization of single qubit transformation, or to build high dimensional quantum states (qudits).

The question of unitary gate synthesis in S-LOQC has already been studied in a series of work over the past few years.
In the seminal paper \cite{lukens2017frequency}, a set-up comprising  2 PS {[P]} and 2 EOM {[P]} in the sequence [PEPE] was first considered.
Using numerical optimization techniques, a deterministic Hadamard gate for spectral dual-rail encoding was theoretically designed with a unity fidelity and success. The question of parallelization was also studied, however it was found that a minimum of 6 modes separating two qubits was needed to ensure a success probability $>90\%$. The number of components, ancilla bits and guard bands required were found to be higher for the synthesis of the $CZ$-gate.\\
In \cite{lu2018electro}, the authors adapted their scheme from \cite{lukens2017frequency} to experimentally implement high-fidelity frequency beamsplitters and tritters with classical light. Two important modifications from their previous proposal involved reducing the number of components from 4 to 3 (in {an [EPE]} sequence), and the use of phase-shifted sinewaves as the EOM RF-driving as opposed to arbitrary waveforms, which place greater demands on bandwidth. Using this {[EPE]-single-tone-} scheme, a beamsplitter was experimentally realized with  fidelity of $\approx 0.99998$ and success probability $\approx 97.39\%$ at 1545.04 nm. The scheme was also shown to be well-suited for parallelization, where the separation between adjacent beamsplitters was found to be a minimum of 4 spectral modes, allowing for 33 beamsplitters to be implemented in parallel under the constraints of the experiment. Furthermore, with the addition of an extra tone to the EOM RF-driving, a frequency tritter with fidelity $\approx 0.9989$ and success probability $ \approx 97.30\%$ was synthesized.\\
The  same {[EPE] } scheme was then used to implement distinct quantum gates (Hadamard and Identity) in parallel, in frequency-bin encoding \cite{lu2018quantum}.  The corresponding programmable unitary was moreover used  to tune the overlap between adjacent spectral bins, which allowed to observe spectral Hong-Ou-Mandel interference with a visibility of 97$\%$. More generally, it was shown that individual single-qubit gate operations can be applied in parallel to each of an ensemble of co-propagating qubits, where each operation can be smoothly tuned between the Identity and Hadamard gates.\\
More recently arbitrary control of spectral qubits was reported in \cite{Lu_2020} with the same set of components [EPE], (i) experimentally with a single-tone sinewave modulation, and (ii) numerically for dual-tone modulation. 
The 2-qubit CNOT gate with the same setup ([EPE] and BFC source) \cite{lu2019controlled} was also reported. Gate reconstruction was performed from measurements in the two-photon computational basis alone, and a fidelity $\mathcal{F} \approx 0.91$ was inferred.


In our work we explore S-LOQC further, with a focus on single qubit unitary gate synthesis, first for a single qubit, and then for many qubits in parallel. 
S-LOQC is indeed naturally adapted for parallelization allowing, without substantial change in the implementation, to apply the same unitary gate to many qubits in parallel. 
While being used in seveval quantum information protocols \cite{samara_high-rate_2019,xiong_compact_2015} time-bin encoding has not been as much explored as frequency {encoding} in the context of the Spectral LOQC platform. It has however been proved {to be} efficient for several tasks in quantum information processing \cite{Lo:18,PhysRevLett.111.150501,fang2018highcapacity,madsen_quantum_2022}. Similarly to frequency encoding, it offers the advantages of a high dimensional accessible Hilbert space and we show here that time encoding allows for more expressive gate synthesis possibilities than frequency encoding  {while} using the same number of phase modulators and pulse-shapers components.

 We investigate both [EPE] and [PEP] configurations of components, and show that both configurations can be used to efficiently perform any unitary transformation, with unity fidelity and success probability. 
 In particular we exhibit an analytical solution for the  synthesis of an arbitrary single-qubit unitary, using minimal ressources, i.e. a single-tone Radio Frequency (RF) driving for the EOMs.
Moreover we also show that those transformations can be applied in parallel with time encoding to a larger number of qubits than what is possible with frequency encoding. We study, for several gates, the trade-off between the quality (fidelity, success probability) of the synthesis, the parallelization that can be achieved, and the ressources needed, in particular in terms of RF bandwidth.\\

The article is organized as follows: we introduce in section II the formalism for encoding both in time or frequency basis and for the corresponding Hilbert spaces, and the description of the components in both basis.
Section III  defines the general problem of qubit unitary gate synthesis that we tackle in this article and explains the rationale of our approach. We  notably detail the  "two-scattering model for the pulse-shaper in the time basis, that will be instrumental for the new results reported in section IV. We finally detail how the performance parameters for gate synthesis, namely fidelity and success probability are defined.
The next two sections report our results, that consist in a systematic and wide exploration of the different options, in terms of encoding bases, gates,  and configurations (either [EPE] or [PEP]).
Section IV presents our results related to the synthesis unitary gates, for a single qubit.
Remarkably, we exhibit, in the time basis, an exact solution allowing the synthesis of an arbitrary single-qubit unitary with single tone RF driving of the $EOM$, in both [EPE] and [PEP]  configurations. These results are put in perspective with single qubit unitary gate synthesis in the frequency basis, whose fidelity and success probability depend on the type of gate that is targeted.
Section V then presents our results related to the parallel synthesis of the same qubit gate over many different qubits. Here again, we compare the two encodings, as well as the two considered configurations, and discuss the interplay between the performance of the synthesis and the number of RF tones.

\section{Time and frequency formalism for S-LOQC}

In this section we define the photonic states that we will employ and the theoretical description of the Pulse Shaper (PS) and the Electro-Optic Phase Modulator (EOM). The descriptions of the devices must take into account the EOM and PS physical characteristics, as these characteristics set a limit on the total number of available modes for quantum manipulation.

\subsection{Optical modes and quantum states}
The system that we consider is composed of $M$ optical modes identified either by a frequency bin of width $\delta\omega$, centered on $\omega_j$ or by a time bin of width $\delta t$, centered on $t_k$. Frequency modes $\ket{\omega_j}$ are linear combinations of time-bin modes $\ket{t_k}$. These two sets of basis vectors are connected by a discrete Fourier transform



\begin{align}
|t_k\rangle &\approx \frac{1}{\sqrt{M}}\sum_{j=0}^{M-1} \exp\left(i \frac{2\pi}{M} jk\right)|\omega_j\rangle,\\
|\omega_j\rangle &\approx \frac{1}{\sqrt{M}}\sum_{k=0}^{M-1} \exp\left(-i \frac{2\pi}{M} jk\right)|t_k\rangle.
\end{align}


The interchange between the frequency and time bases is exact only in the continuous case, when $M\rightarrow\infty$ and $\delta\omega,\delta t\rightarrow0$. 

To operate on single qubits, we divide our Hilbert space of dimension $M$ into $M/2$ independent subspaces according to a choice of qubit encoding either in the frequency domain $\mathcal{H}_j^{\omega}$ using two contiguous frequency bins or in the time domain $\mathcal{H}_k^{t}$ using modes separated by $M/2$ time bins, with $j,k \in [0, M/2-1]$. 

\begin{align}
    \mathcal{H}_j^\omega = \left\{\ket{\omega_{2j}},\ket{\omega_{2j+1}}  
    \right\} 
    ,\,\omega_{2j+1}-\omega_{2j}=\delta\omega,\label{definition_encoding1}\\
    \mathcal{H}_k^t = \left\{\ket{t_{k}},\ket{t_{k+M/2}}  \right\} 
    ,\,t_{k+M/2}-t_k=\frac{M}{2}\delta t,\label{definition_encoding2}
\end{align}
Such a qubit encoding choice is made to take into account the technical limitations of the devices. In the present case, the fact that the EOM couples mainly neighboring frequency modes justifies the choice of adjacent modes to encode frequency-bin qubits. Conversely, the fact that the PS couples time modes that are M/2 time bins appart (cf Appendix \ref{PS_temp_basis} ) justifies the choice of encoding used for the time-bin qubits.\\

As the considered Hilbert spaces are orthogonal, we can write the sum of the sub-spaces as
\begin{align}
\mathcal H = \bigoplus_j\mathcal{H}_j^\omega = \bigoplus_k \mathcal{H}_k^t.
\end{align}
We now introduce the devices that we are using for photonic qubit processing.

\subsection{Pulse Shaper} 
\label{PShaper}
The phase only programmable filters (PS) can shift the phase of frequency modes concurrently and independently. It is therefore characterized by a set of $M$ angles $\{\varphi_j\}$ (one angle per mode) that can be chosen freely, without constraints. In general, the PS acts as the tensor product of distinct phase shifts operators, one per frequency mode and its operator $\hat{U}_{PS}$ can be written

\begin{align}
\hat{U}_{PS} = \bigotimes_{j=0}^{M-1} \exp(i \varphi_j \hat{N}_{\omega_j}),
\end{align}
where $\hat{N}_{\omega_j}$ is the number operator of the $\omega_j$ frequency mode: $\hat{N}_{\omega_j} = \sum_n n |n_{\omega_j}\rangle\langle n_{\omega_j}|$.
However in the restricted case of a single photon, we can thus treat the PS as a diagonal matrix P in the frequency basis. Its action on the 1-photon subspace is described in Fig. \ref{schema_devices}.b.
\begin{align}
P|\omega_j\rangle = e^{i\varphi_j}|\omega_j\rangle.
\end{align}

\begin{figure}
    \begin{center}
        \includegraphics[width=0.5\textwidth]{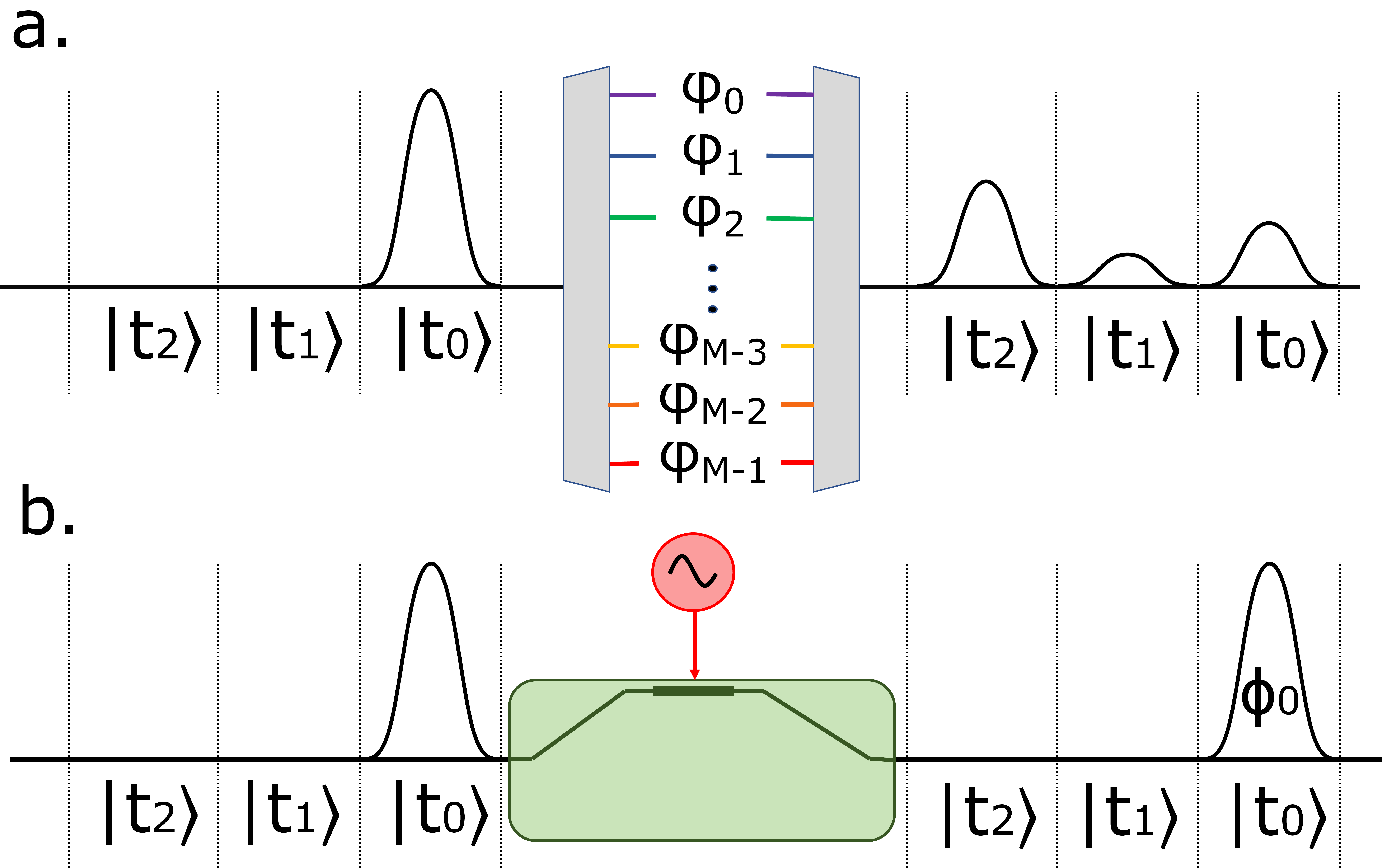}
        \caption{Action of the Pulse-shaper (a.) and Electro Optic Phase Modulator (b.) in the time basis.}
        \label{EOM_PS_time}
    \end{center}
\end{figure}

In order to describe the action of the PS in the time basis (see Fig. \ref{EOM_PS_time}), we use the Fourier transform  $\tilde{P} = FPF^\dagger$ where the matrix $F$ and the detailed calculations are given in appendix \ref{PS_temp} and we obtain

\begin{align}
\tilde{P}|t_k\rangle = \frac{1}{M}\sum_{k'=0}^{M-1}\sum_{j=0}^{M-1} \exp \left(i\frac{2\pi}{M}(k'-k)j+\varphi_j\right)\ket{t_{k'}}.
\end{align}

The PS is therefore a scatterer in the time basis, coupling a single time mode to all the other modes. The number of modes over which a single mode can be scattered will be an important optimization parameter, as shown in section \ref{stability}.\\

\subsection{Electro-Optic Phase Modulator (EOM)}
The EOM is a device that is complementary to the PS as it performs phase shifts of individual modes in the time domain, and therefore its operator can be expressed as 

\begin{align}
\hat{U}_{EOM} = \bigotimes_{k=0}^{M-1} \exp(i \phi_k \hat{N}_{t_k}),
\end{align}
where $\hat{N}_{t_k}$ is the number operator of the $t_k$ time-bin mode: $\hat{N}_{t_k} = \sum_n n |n_{t_k}\rangle\langle n_{t_k}|$.
Within our restriction to the 1-photon subspace, we can write the action of the EOM over time modes $\ket{t_k}$ as a diagonal matrix $E$ in the $\{|t_k\rangle\}$ basis as
\begin{align}
E\ket{t_k} = e^{i\phi_k}\ket{t_k}.
\end{align}
\begin{figure}
\begin{center}
    \includegraphics[width=0.5\textwidth]{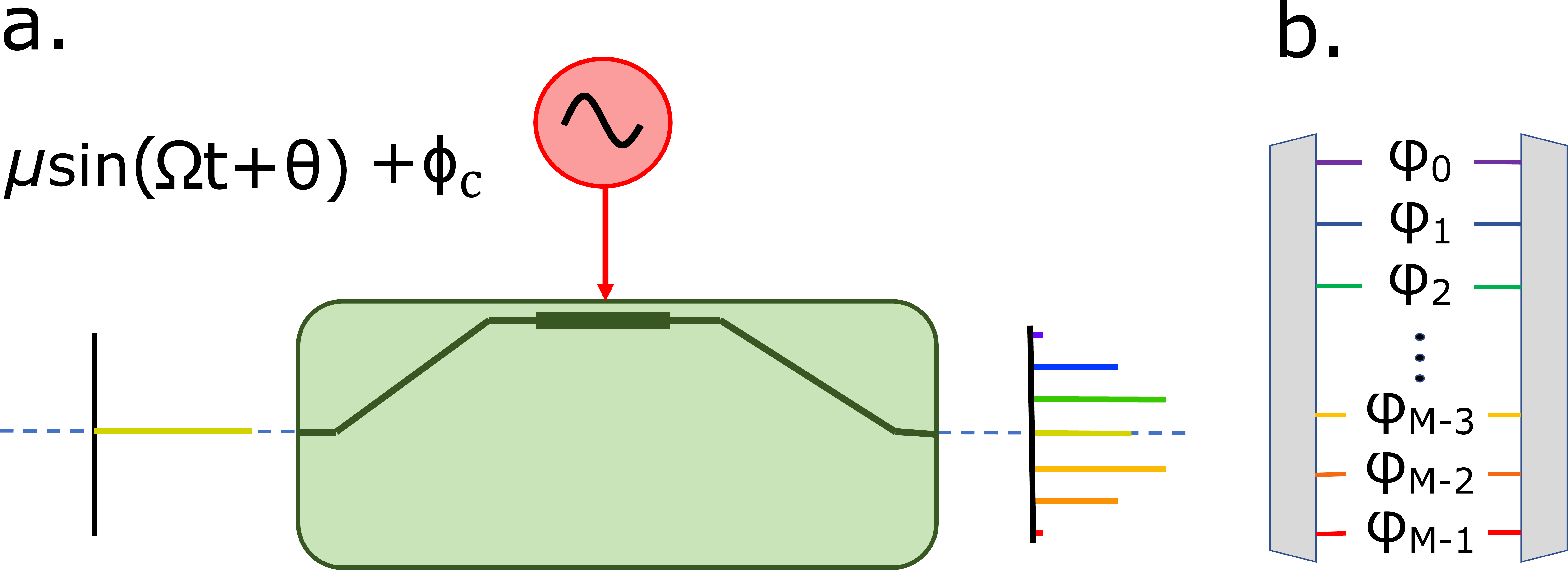}
    \caption{Action of a. the Electro-optic Phase Modulator and b. the Pulse Shaper in the frequency basis}
    \label{schema_devices}
\end{center}
\end{figure}
Contrary to the PS case, the set of angles $\{\phi_k\}$ for the EOM cannot be chosen freely. It is determined by the RF driving.
For instance, in the case of a monochromatic (or single tone) RF driving, the phase $\phi_k$ applied to the optical wave can be written as a sinusoidal function of time with frequency $\Omega$
\begin{align}
\phi_k = \mu\sin(\Omega t_k + \theta) + \phi_c,
\end{align}\label{phik}
with $\mu = \pi\frac{V_m}{V_\pi}$ the modulation index proportional to the modulation amplitude $V_m$ and $V_\pi$ the half-wave voltage of the EOM, $\theta \in [0,\pi]$ a constant angle 
and $\phi_c\in [0,\pi]$ a constant phase shift applied to all time bins.
In practice, $\mu$ is limited by the power of the RF source that drives the EOM (typically $\mu$ is of order 1), and by the EOM characteristics. The maximum frequency $\Omega$ is determined by the frequency response of the EOM. 

We can describe the action of the EOM in the frequency basis \cite{olislager2014creating,capmany2011quantum} (see Fig. \ref{schema_devices}) through the Fourier transform $\tilde {E} = FEF^\dagger$ and check that the EOM creates left and right side bands (spaced by $\Omega$) with amplitude decreasing according to Bessel functions. 
If the mode spacing $\delta \omega$ is set equal to the RF driving single-tone $\Omega$, we have
\begin{equation}\label{EOM}
\tilde{E}\ket{\omega_j} = e^{i\phi_c}\sum_{k = -\ceil*{\mu}-1}^{k = \ceil*{\mu}+1}(e^{i\theta})^{k}J_{k}(\mu)\ket{\omega_{j+k}}.
\end{equation}

The sum can be extended to $\pm\infty$ as the additional terms are weighed by vanishing Bessel functions \cite{tsang2011cavity}. 

The EOM is therefore a scatterer in the frequency basis.
\subsection{Orders of magnitude}
\label{subsec:oom}
\color{black}
The PS and the EOM act on conjugate variables (frequency and time), which are related by a Fourier transform, which results in relationships between bandwidths and resolutions. We denominate $\Delta \omega$ and $\delta \omega$ respectively the total frequency bandwidth that the PS can operate on and its frequency resolution. On the other hand, if the EOM is driven at a RF frequency $\Omega$, it can generate sidebands at intervals $\Omega$. This naturally defines a number $M_\omega = \Delta \omega/\Omega$ frequency modes as long as $\delta \omega \lesssim \Omega$, so that the PS can distinguish and address them individually. 
It is important to match the number of frequency modes and time modes to make sure that the two devices are acting on two conjugate bases of the same Hilbert space.

In the time domain, the resolution $\delta t$ will be limited by the characteristic times of the detectors. The period of the RF frequency driving the EOM is $T= 2\pi/\Omega$, which in principle defines $M_t = T/\delta t = \Delta \omega/\Omega=M_\omega$ time modes. 

A realistic PS ( Finisar WaveShaper 4000A) has a total bandwidth $\Delta\omega$ = 5.36 THz, and the RF voltage signal that drives the EOM can be set to be a sine wave with a frequency in the order of 10 GHz (which is close to the frequency resolution of the PS). From these figures we derived a number of time and frequency modes $M = T / \delta t  = \Delta \omega / \delta \omega\approx $536 and in our calculations, we chose $M=2^7=128$ modes.

\section{Problem Definition: qubit gate synthesis with EOM and PS }

\subsection{Objective}

Our aim is to synthesize single-qubit quantum gates using some combination of  the components described in the previous section namely pulse-shapers  (PS)  and electro-optic modulators (EOM).
A single qubit gate is by definition a unitary transformation over a 2-dimensional Hilbert space, and is hence represented by a 2$\times$2 unitary matrix whereas the components can be described by unitaries ( $P$, $\tilde{P}$, $E$ and $\tilde{E}$) that can operate over a much larger, M-dimensional Hilbert space. 
Describing how to perform single qubit gate synthesis in this context therefore implies to specify several types of information
\begin{itemize}
\item The configuration, i.e. how many components are combined into the unitary. After justifying that the minimum number of components is 3, we focus in this article on two main configurations, [EPE] and [PEP].
\item The values of the different components free parameters, (phases of the form $\varphi_0,\hdots,\varphi_{M-1}$ for the pulse-shaper, and parameters $\mu,\theta,\phi_c$ for an electro-optic modulator driven with a single frequency modulation at $\Omega$).
\item The choice of encoding, i.e. a choice of two orthonormal M-dimensional vectors,  $\{ \ket{e_1},  \ket{e_2}\}$, that constitute the encoding basis and are identified with the logical qubit basis $\{ \ket{0_L}, \ket{1_L}   \}$.
\end{itemize}

As explained in the previous section, we will use both frequency basis $\{ \ket{ \omega_i}, \ket{\omega_j} \}$ and time basis $\{  \ket{t_i}, \ket{t_j} \}$ as defined in equations (\ref{definition_encoding1}) and (\ref{definition_encoding2}).

\subsection{Performance metrics}

For a given configuration (e.g. combination of PS and EOMs), we can define a global operator $\hat{V}(\Phi)$ and the corresponding M$\times$M unitary matrix $V(\Phi)$, product of $n$ PS and $n'$ EOMs, depending on many parameters that we will globally denote as $\Phi \equiv (\varphi_{1,0},\hdots,\varphi_{n,M-1},\mu_{1,0}, \theta_{1,0}, {\phi_c}_{1,0} \hdots \mu_{n',M-1}, \theta_{n',M-1},{\phi_c}_{n',M-1} )$.\\ The choice of the encoding $\{ \ket{e_1},  \ket{e_2}\} \leftrightarrow \{ \ket{0_L}, \ket{1_L}   \}$ then directly induces a reduced 2$\times$2 matrix $W(\Phi)$ defined as

\begin{equation} \label{eq:13}
W(\Phi) = \begin{pmatrix}
\bra{0_L}\bm\hat{V}(\Phi)\ket{0_L} & \bra{0_L}\bm\hat{V}(\Phi)\ket{1_L}\\
\bra{1_L}\bm\hat{V}(\Phi)\ket{0_L} & \bra{1_L}\bm\hat{V}(\Phi)\ket{1_L}
\end{pmatrix}.
\end{equation}

If $T$ represents the (ideal) target qubit unitary that one wants to synthesize, then the performance of the synthesis can be quantified  by comparing this resultant 2x2 matrix  $W(\phi)$ to the target (ideal) gate $T$. The performance of the synthesis can be essentially captured by two parameters: 
\begin{itemize}
\item The success probability $\mathcal{P}$, that measures the stability of the 2-dimensional subspace  $\{ \ket{e_1},  \ket{e_2}\}$, under the unitary $V(\phi)$
\begin{equation} \label{eq:Sproba}
\mathcal{P}(W,T) = \frac {Tr(W^\dagger W)}{Tr(T^\dagger T)},
\end{equation}
\item The fidelity  $\mathcal{F}$, that measures the accuracy of the synthesis of the target single-qubit gate (in other words, it measures how close we are to the gate we intended to synthesize in the first place)
\begin{equation} \label{eq:fidel}
\mathcal{F}(W,T) = \frac {Tr(W^\dagger T)Tr(T^\dagger W)} {Tr(W^\dagger W)Tr(T^\dagger T)}. 
\end{equation}
\end{itemize}

\subsection{Arbitrary unitary synthesis for a single qubit}

We will identify configurations of components, and encoding methods for which it is possible to synthesize {\bf any} single qubit unitary, by varying the free parameters of $\Phi$, and this with fidelity and success probability greater than some given threshold values, that we denote  $\mathcal{F}_{th}$ and  $\mathcal{P}_{th}$. This objective is captured in the definition

{\it{For a given encoding  $\{ \ket{e_1},  \ket{e_2}\} \leftrightarrow \{ \ket{0_L}, \ket{1_L}   \}$, a configuration $\hat{V}(\Phi)$ can perform arbitrary single-qubit unitary synthesis with precision $\mathcal{F}_{th}$,   $\mathcal{P}_{th}$ iff\\$ \forall \, U \in SU(2),\, \exists \, \Phi$ such that  $\mathcal{P}(W(\Phi), U)   \ge  \mathcal{P}_{th}$ and $\mathcal{F}(W(\Phi),U) \ge  \mathcal{F}_{th}$.}}

An arbitrary unitary matrix can be described with four parameters
\begin{equation}\label{arbitrary_unitary}
    \mathcal{M}(a,b,c,d)=e^{ia}
    \begin{pmatrix}
         \cos(\frac{c}{2})e^{-i\left(\frac{b}{2}+\frac{d}{2}\right)}
         & -\sin(\frac{c}{2})e^{-i\left(\frac{b}{2}-\frac{d}{2}\right)}  
         \\
         \sin(\frac{c}{2})e^{i\left(\frac{b}{2}-\frac{d}{2}\right)}
         & \cos(\frac{c}{2})e^{i\left(\frac{b}{2}+\frac{d}{2}\right)} 
    \end{pmatrix}
\end{equation}
\color{black} 
where $a,b,c,d\,\in\, \mathbb{R}$. Note that if an arbitrary single-qubit unitary synthesis is achieved with $\mathcal{F}=1$ and $\mathcal{P}=1$ with a given configuration and some encoding, then any other basis choice within the subspace  $\{ \ket{e_1},  \ket{e_2}\} $ will lead to an arbitrary single qubit synthesis as the change of encoding is a unitary operation.
On the other hand, if a configuration cannot achieve arbitrary synthesis, then the ability to synthesize one given target gate $U$ will vary with the basis choice.

\subsubsection{Subspace stability and choice of encoding\label{stability}}
We aim in general at the highest possible $\mathcal{F}$ and $\mathcal{P}$ for every gate synthesis.
The requirement on the success probability $\mathcal{P}$ can be used as a guideline, for the choice of the encoding basis we consider. A necessary condition to achieve a high success probability is to pick an encoding subspace that is (approximately) stable under the action  of $\hat{V}(\phi)$. 
From the previous considerations on the pulse-shaper and the electro-optic phase modulator, we get a first intuition on the difficulties of obtaining some transformations, according to the considered encoding :
some require energy exchange between the two basis states as for instance the X or Y Pauli transformations whereas some only require phase changes as the Z Pauli transformation. A very important example is that of the Hadamard transformation that achieves the balanced splitting of the energy between the two modes corresponding to the basis states. It is therefore important to choose the component combination and encoding in order to be able to control both phase changes and energy exchanges between the two modes of the computational basis.
In any component combination, we need either a dephasing element, a scatterer, or a certain combination of these two elements. The stability condition will be satisfied if the combination of components and the encoding allow the scattered energy to remain in the subspace defining the qubit.
Let us now consider the consequence of this requirement, for our two families of encoding.

\paragraph{Frequency encoding}
In the frequency basis, all vectors are stable under the action of the pulse shaper but the EOM acts as a scatterer. The number of modes where the energy is scattered from one given initial mode depends on the modulation index. As mentioned, because the EOM couples frequency modes locally (see Eq. (\ref{EOM}) and Fig. \ref{schema_devices}), the computational basis is formed of adjacent modes, of the form $\{ \ket{ \omega_i}, \ket{\omega_i+1} \}$  when studying synthesis in the frequency encoding as in \cite{lukens2017frequency}. 
 
\paragraph{Time encoding}
In the time basis, all vectors are stable under the action of the EOM but the PS acts as a scatterer. The large number of independent parameters available for the PS allows a variety of scattering possibilities.

\begin{center}
\fbox{\begin{minipage}{0.9\textwidth}
\begin{proposition}\label{2scat}
PS parameters can be chosen to satisfy the two-scattering assumption
\begin{equation*}
\forall k\in  [0, M/2-1] , \qquad    \tilde{P}\ket{t_k}=\alpha\ket{t_k}+\beta\ket{t_{k+M/2}},
\end{equation*}
where $\alpha$ and $\beta$ are complex numbers, linked by the unitarity condition $|\alpha|^{2}+|\beta|^{2}=1 $.
This ensures the stability of our time qubit subspace in all transformations including PS and EOM when using the basis choice $\{\ket{0_L}= \ket{t_k}, \ket{1_L}=\ket{t_{k+M/2}} \}$.
\end{proposition}
\end{minipage}}
\end{center}


 
In Appendix \ref{PS_temp_basis} we calculate the PS parameters that allow to satisfy the two-scattering assumption and we derive the matrix elements of the PS both in the frequency basis and in the time basis

\begin{equation}
    P\ket{\omega_j}=\left[\alpha+\beta e^{-i 2 \pi m\frac{j}{M}}\right]\ket{\omega_j}.
\end{equation}

We obtain
\begin{equation}
\left\{
\begin{array}{c}
    \forall k\in \llbracket 0, \frac{M}{2}-1 \rrbracket, \tilde{P}\ket{t_k}=\alpha\ket{t_k}+\beta\ket{t_{k+M/2}}\\
    \forall k\in \llbracket \frac{M}{2}-1, M-1 \rrbracket, \tilde{P}\ket{t_k}=\beta\ket{t_k}+\alpha\ket{t_{k-M/2}}.
\end{array}
\right.
\end{equation}

The matrix in the time basis is therefore
\begin{equation}\label{PS_temp_matrix}
    \tilde{P} =  e^{i\gamma}
    \left( \begin{array}{ccc}
    |\alpha|I_{\frac{M}{2}} & \pm i|\beta|I_{\frac{M}{2}} \\
    \pm i|\beta|I_{\frac{M}{2}} & |\alpha|I_{\frac{M}{2}} 
    \end{array} \right),
\end{equation}
where $I_{\frac{M}{2}}$ stands for a Identity matrix of dimension M/2 and $\gamma$ is a global phase induced by the pulse shaper.
When using the basis choice $\{\ket{0_L}= \ket{t_k}, \ket{1_L}=\ket{t_{k+M/2}} \}$, the $2\times 2$ matrix of the PS can be mapped to the matrix of a rotation of angle $\theta_{PS}$ about the $x$-axis in the Bloch sphere, with:

\begin{equation}
    \begin{cases}
e^{i\gamma}  |\alpha|=\cos{\frac{\theta_{PS}}{2}};\\
e^{i\gamma}|\beta|=\sin{\frac{\theta_{PS}}{2}}
    \end{cases}.
\end{equation}
It can also be mapped to the matrix of an arbitrary beamsplitter with reflexion and transmission coefficients $r$ and $t$

\begin{equation}
    \begin{cases}
    |\alpha|=|r|\\
    |\beta|=|t|
    \end{cases}.
\end{equation}

In the same way, the matrix of the EOM in the time basis can be written:
\begin{equation}
    E = e^{i\phi_c}
    \left( \begin{array}{ccc}
    diag\{e^{\phi_k}\}_{0 \leq k \leq \frac{M}{2}-1} & 0 \\
    0 & diag\{e^{-\phi_k}\}_{0 \leq k \leq \frac{M}{2}-1}
    \end{array} \right).
\end{equation}
With the chosen encoding, the $2\times 2$ matrix of the EOM can be identified to the matrix of a rotation of angle $\theta_{EOM}$ about the $z$-axis in the Bloch sphere, with:
\begin{equation}
    \begin{cases}
   i\phi_c+\phi_k  =-i\theta_{\text{EOM}}/2\\
    i\phi_c-\phi_k  =i\theta_{\text{EOM}}/2
    \end{cases}.
\end{equation}

\subsubsection{Minimum number of components}


Using the analogy with rotations and the fact that an arbitrary unitary transformation requires the composition of three rotations about two orthogonal axes of the Bloch sphere \cite{nielsen2000quantum}, we derive that a minimum number of three components will be required to synthesize an arbitrary unitary transformation, using pulse-shapers and electro-optic phase modulators. Since two EOMs or two PSs in serial are equivalent to one, 
the sequences with minimum resource to investigate are [EPE] and [PEP].
We investigate the case of the single qubit for both frequency and time bases encoding.
\color{black}

\section{Results for single qubit gate synthesis}

\begin{table}
    \centering
    \includegraphics[width=0.5\textwidth]{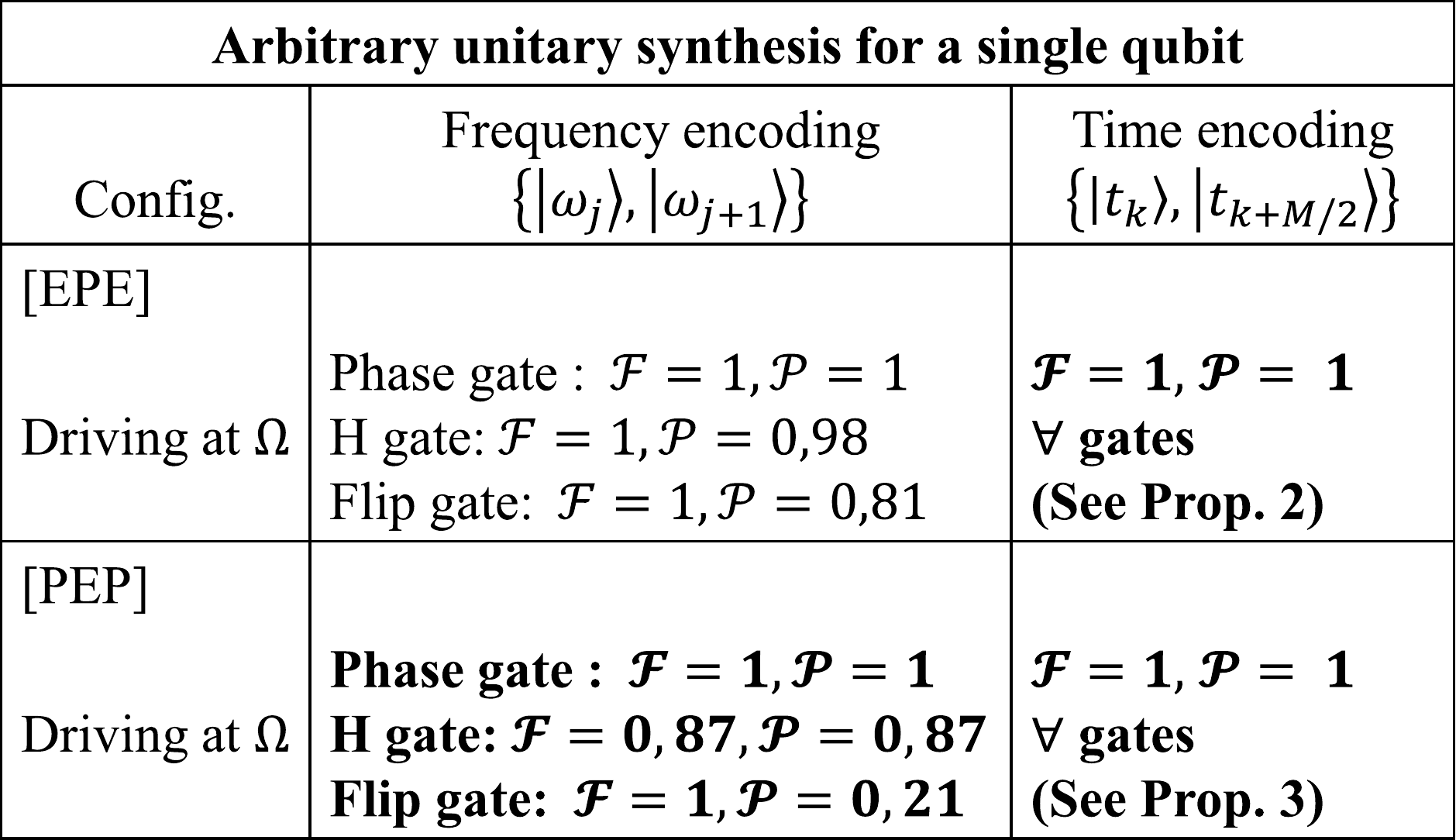}
    \caption{Comparison of the 
   performances of single qubit arbitrary unitary synthesis with [EPE] and [PEP] configurations  for single tone driving  at $\Omega$ of the EOMs and for two encodings. \\The bold part corresponds to  {\bf  new results reported in this work}}\label{AUSSQ}
\end{table}
\normalsize
\subsection{Frequency encoding}
Within each frequency subspace, we define the logical basis as

\begin{equation}\label{fr}
\begin{array}{cc}
    \ket{0_L}_j=\ket{\omega_j}\qquad
    \ket{1_L}_j=\ket{\omega_{j+1}}.
\end{array}{}
\end{equation}

\subsubsection{[EPE] configuration}
This configuration has been thoroughly investigated \cite{lukens2017frequency,lu2018quantum,lu2018electro, lu2019controlled,Lu_2020} and the  theoretical results about one-qubit gates  synthesis are summarized in Table \ref{AUSSQ}.
Experimental implementation of Hadamard have been reported in \cite{lu2018electro}, the control-not gate in \cite{lu2019controlled} and more recently arbitrary transformations of a single qubit in \cite{Lu_2020}.
The limitation on success probability arises from the difficulty to control the scattering of energy in many modes by the EOM. This is also the reason why guard-bands must be used to avoid cross talk when working with qubits in parallel in the frequency domain : the coupling to adjacent modes cannot be perfectly controlled, even with ideal lossless devices.

\subsubsection{[PEP] configuration}

In order to assess this configuration, the first step is to write the generic element of the corresponding synthesized matrix, using two diagonal matrices $P_1$ and $P_2$ for the PS and Eq.(\ref{EOM}) for the EOM matrix $\tilde{E}$ in the frequency basis
\begin{equation}
    \forall \, k,k'\in \left[ 0,M-1\right], \, (P_2\tilde{E}P_1)_{k'k} = e^{i(\varphi_{1,k}+\varphi_{2,k'}+\phi_c)}\sum_{k = k'-\ceil*{\mu}-1}^{k =k'+ \ceil*{\mu}+1}(e^{i\theta})^{k'-k}J_{k}(\mu).
\end{equation}

For qubit $j$, $\forall j \in  [0, M/2-1] $ encoded over modes $\ket{\omega_j}$ and $\ket{\omega_{j+1}}$, the matrix of the synthesized unitary is given by
\begin{equation}\label{W_PEP}
    \mathbf{W_j} =  e^{i\phi_c}
    \left( \begin{array}{ccc}
    J_0{(\mu)}e^{i(\varphi_{1,2j}+\varphi_{2,2j})} 
    & e^{-i\theta}J_1{(\mu)}e^{i(\varphi_{1,2j+1}+\varphi_{2,2j})}  \\
  - e^{i\theta} J_1{(\mu)}e^{i(\varphi_{1,2j}+\varphi_{2,2j+1})}  
  & J_0{(\mu)}e^{i(\varphi_{1,2j+1}+\varphi_{2,2j+1})}  
    \end{array} \right),
\end{equation}
where $\varphi_{n,m}$ is the phase applied by PS$_n$ to mode $m$.

Let $\mathcal{M}(a,b,c,d)$ be the matrix of an arbitrary unitary as given in Eq. \ref{arbitrary_unitary}. One can achieve an exact synthesis of U over qubit j \textbf{iff} parameters $\varphi$ match the following set of conditions:
\begin{equation}
    \begin{cases}
        J_0{(\mu)} = \cos{(c/2)} \\
        J_1{(\mu)} = \sin{(c/2)} 
    \end{cases}
    \begin{cases}
        \varphi_{1,2j} =  \frac{-d -\theta + \pi + s_j}{2}  \mod{\pi} \\
        \varphi_{2,2j} =  -\frac{b+d}{2}+ \varphi_{1,2j} \mod{\pi} \\
        \varphi_{1,2j+1} = s_j - \varphi_{1,2j} \mod{\pi} \\
        \varphi_{2,2j+1} = -s_j +\big( \frac{b - d}{2} + \varphi_{1,2j} \big) \mod{\pi}.
    \end{cases}
\end{equation}

\begin{figure}
\begin{center}
    \includegraphics[width=0.5\textwidth]{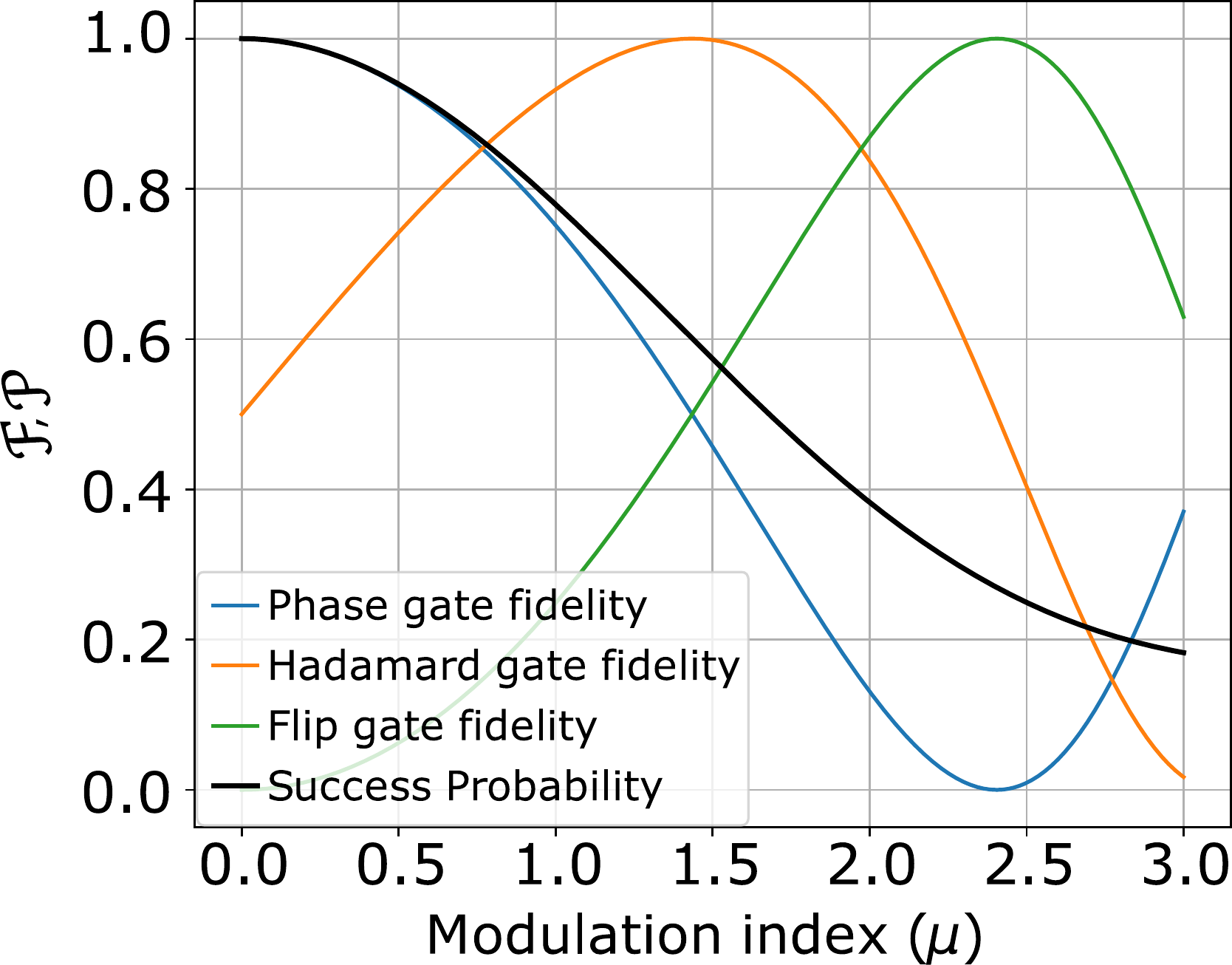}
    \caption{Success probability $\mathcal{P}$ and fidelity $\mathcal{F}$ for phase, Hadamard and X gates in the [PEP] configuration for frequency-bin qubits. Varying $\mu$ covers all angles $\theta$ of the Bloch sphere following Eq. \ref{W_PEP} and the performance do not depend on the rotation around the $z$ axis.
    Black solid line: success probability, standing for all gates, as it represents the amount of energy remaining in the computational space; blue solid line: fidelity of the phase gate synthesis; green solid line: fidelity of the Hadamard gate synthesis; red solid line: fidelity of the X-gate synthesis;}
    \label{pep_freq_sp_fid}
\end{center}
\end{figure}

If the parameters of the pulse shapers satisfy the second set of equations, then fidelity and success probability do not depend on the qubit number $j$ and are given by
\begin{equation}
\mathcal{P} = J_0{(\mu)}^2 + J_1{(\mu)}^2,
\end{equation}
\begin{equation}
\mathcal{F} = \frac{ \big[ J_0{(\mu)}\cos{(\frac{c}{2})} 
+J_1{(\mu)}\sin{(\frac{c}{2})} \big]^{2} }
{J_0{(\mu)}^2 + J_1{(\mu)}^2}.
\end{equation}
Subsequently we get the following results on the best achievable performances of the [PEP] configuration, indcating that 
 unit success probability and unit fidelity can only be obtained for phase gates.

\begin{equation}
    \begin{cases}
        \mathcal{P}=1 \Leftrightarrow \mu = 0 \\
        \mathcal{F}=1 \Leftrightarrow J_0(\mu)\sin(\frac{c}{2}) = J_1(\mu)\cos(\frac{c}{2}) \\
        \mathcal{P}=\mathcal{F}=1 \Leftrightarrow \mu = c = 0.
    \end{cases}
\end{equation}
Figure \ref{pep_freq_sp_fid}, summarizes the performance of gate synthesis for this configuration. Varying $\mu$ allows to cover all unitaries. The dephasing applied to each frequency mode by the pulse shapers are chosen such that $\mathcal{F}=1$ is achievable for all gates but with a decreasing success probability $\mathcal{P}$.  The Bit Flip gate $X$ appears,  for this configuration, as the worst case in terms of gate synthesis objective: it is  the gate for which the success probability,  conditioned on unit fidelity, is the lowest.

We can see some intrinsic limitation of the frequency encoding, when using the [PEP] configuration, due to the loss of energy in the side bands induced by the EOM. 
The gates requiring no coupling or energy exchange between the two modes of the qubit (phase gates) exhibit unity success probability, as all the energy remains in the computational modes. For the other gates however, the more coupling we need, the greater the index of modulation, and the more energy gets lost in the adjacent modes, leading to a lower success probability. In summary, the success probability in this case is lower than in the previous configuration, especially for the X gate: with only one EOM, 
the energy scattered in adjacent modes cannot be coupled back to the qubit subspace.
\color{black}
While this \NB{[PEP]} configuration had been studied with spectral encoding for quantum state tomography and probabilistic Hong-Ou-Mandel \cite{imany_50-ghz-spaced_2018,imany_frequency-domain_2018,kues_-chip_2017,khodadad_kashi_spectral_2021}, it 
it had not been explored, up to now, for arbitrary gate synthesis.


\subsection{Time Encoding}\label{TE}
 In this basis, the matrices of the two components are
 \begin{equation}
    \!\!\!\tilde{P} =  e^{i\gamma}
    \left( \begin{array}{cc}
    |\alpha|I & \pm i|\beta|I \\
    \pm i|\beta|I & |\alpha|I 
    \end{array} \right);\,\,
    E = e^{i\phi_c}
    \left( \begin{array}{cc}
    diag\{e^{\phi_k}\}_{0 \leq k \leq \frac{M}{2}-1} & 0 \\
    0 & diag\{e^{-\phi_k}\}_{0 \leq k \leq \frac{M}{2}-1}
    \end{array} \right),
\end{equation}
The qubit is encoded over two time modes separated by $M/2$ modes
\begin{equation}\label{temp}
\begin{array}{cc}
    \ket{0_L}_k=\ket{t_k} \qquad
    \ket{1_L}_k=\ket{t_{k+M/2}}.
\end{array}
\end{equation}

\subsubsection{[EPE] configuration}
\begin{center}
\fbox{\begin{minipage}{0.9\textwidth}
\begin{proposition}\label{EPE_temp_single_qubit}
One can achieve exact synthesis of any single qubit arbitrary unitary $U$ with a single frequency RF driving with the [EPE] configuration.
\end{proposition}
\end{minipage}}
\end{center}
\begin{proof}
Let $\alpha$, $\beta$ and $\gamma$ be the parameters of the pulse shaper in the time basis, $\mu_i$, $\theta_i$ the parameters of the $EOM_i$ ($i \in \{1,2\}$) and $\mathcal{M}(a,b,c,d)$ the matrix of an arbitrary unitary $U$. We consider qubit $k \in [0, M/2-1]$ encoded on modes $\ket{t_k}$ and $\ket{t_{k+M/2}}$. 
The matrix of the three-component configuration is
\begin{equation}
    V = E_2\tilde{P}E_1 = e^{i(\phi_{c,1} +\phi_{c,2} + \gamma)}
    \left( \begin{array}{ccc}
    |\alpha|diag\{e^{i(\phi_{1,k}+\phi_{2,k})}\}
    &   \pm i|\beta|diag\{e^{i(-\phi_{1,k}+\phi_{2,k})}\}  \\
      \pm i|\beta|diag\{e^{i(\phi_{1,k}-\phi_{2,k})}\} 
    & |\alpha|diag\{e^{-i(\phi_{1,k}+\phi_{2,k})}\}
    \end{array} \right),
\end{equation}
which yields the matrix of the synthesized unitary for the qubit encoded over modes $\ket{t_k}$ and $\ket{t_{k+M/2}}$
\begin{equation}\label{W_epe}
\forall k \in  [0, M/2-1],
    \mathbf{W_k} =  e^{i(\phi_{c,1} +\phi_{c,2} + \gamma)}
    \left( \begin{array}{ccc}
   |\alpha|e^{i(\phi_{1,k}+\phi_{2,k})}
    &  \pm i|\beta|e^{i(-\phi_{1,k}+\phi_{2,k})}  \\
   \pm i|\beta|e^{i(\phi_{1,k}-\phi_{2,k})} 
  & |\alpha|e^{-i(\phi_{1,k}+\phi_{2,k})}
    \end{array} \right) .
\end{equation}
Identification with $\mathcal{M}(a,b,c,d)$ yields the following set of conditions for qubit $k$
\begin{equation}
	\left\{ \begin{array}{cc}
        |\alpha| = \cos{(c/2)} \\
        |\beta| = \sin{(c/2)} 
    \end{array} \right. \\
    \left\{ \begin{array}{cc}
        \phi_{1,k} + \phi_{2,k} =  \frac{-b-d}{2} \mod{2\pi} \\
        -\phi_{1,k} + \phi_{2,k} + \pi/2 =    
        \frac{-b+d}{2}+\pi \mod{2\pi} \\
        \phi_{1,k} - \phi_{2,k} + \pi/2 = 
        \frac{b-d}{2} \mod{2\pi} \\
        -\phi_{1,k} - \phi_{2,k} =
        \frac{b+d}{2} \mod{2\pi}, \\
    \end{array} \right.
\end{equation}
which finally gives
\begin{equation}
	\left\{ \begin{array}{cc}
        |\alpha| = \cos{(c/2)};\qquad |\beta| = \sin{(c/2)}\\
        \phi_{1,k} =  \frac{- d}{2} - \pi/4  \mod{\pi} \\
        \phi_{2,k} =  \frac{- b}{2} + \pi/4  \mod{\pi}.
    \end{array} \right. 
\end{equation}
As we consider a sinusoidal RF driving,  the dephasing applied by the EOM to each temporal mode $\ket{t_k}$ is $\phi_k = \mu \sin{(\frac{2k\pi}{M}+\theta)}$, ($\forall k \in \llbracket 0, M/2-1 \rrbracket$), which gives
\begin{equation}
    \left\{ \begin{array}{cc}
        |\alpha| = \cos{(c/2)};\qquad |\beta| = \sin{(c/2)};\qquad\mbox{ with } c\in [0,\pi]\\
        \phi_{1,k} =\mu_1 \sin{(\frac{2k\pi}{M}+\theta_1)} 
        = -\frac{d}{2} - \frac{\pi}{4} \mod{\pi} \\
       \phi_{2,k}= \mu_2 \sin{(\frac{2k\pi}{M}+\theta_2)} 
        = -\frac{b}{2} + \frac{\pi}{4} \mod{\pi}.
    \end{array} \right.
\end{equation}
    The condition is satisfied with the following parameters
\begin{align}
	        \mu_1 = \frac{\frac{- d}{2} - \frac{\pi}{4}}
        {\sin{(\frac{2k\pi}{M}+\theta_1)}}&; \qquad 
        \mu_2 = \frac{\frac{- b}{2} + \frac{\pi}{4}}
        {\sin{(\frac{2k\pi}{M}+\theta_2)}} \\
        \theta_1 ,\theta_2 &\ne -\frac{2k\pi}{M}.\nonumber
\end{align}
\end{proof}
This possibility of synthesizing an arbitrary unitary for a single qubit with unit success probability and fidelity is a noticeable advantage of the time encoding as compared to the frequency encoding for which only phase gates can exhibit such performance.

\subsubsection{[PEP] configuration}
\begin{center}
\fbox{\begin{minipage}{0.9\textwidth}
\begin{proposition}\label{PEP_temp_single_qubit}
One can achieve an exact synthesis of an arbitrary unitary $U$ over qubit k with a single frequency RF driving with the [PEP]  configuration.
\end{proposition}
\end{minipage}}
\end{center}
\begin{proof}
Based on the two-scattering property demonstrated in section \ref{stability}, we can see the PS acts as a beamsplitter in the time basis. The following EOM can then introduce an arbitrary phase shift  between the \NB{two modes} of the qubit subspace while the final PS of this configuration acts as a second arbitrary beamsplitter, completing a Mach Zehnder interferometric device. Such Mach-Zehnder has been shown to achieve arbitrary unitary qubit synthesis \cite{englert2001universal}.
\end{proof}
\subsubsection{Trivial gates}
It is noticeable that some gates do not require the three component configuration. For instance, in the time basis, phase gates only require one EOM and no PS. X gates that do not even require any EOM can be trivially obtained in both [EPE] and [PEP] configurations. They correspond to the parameters $b=-d=\frac{\pi}{2}$ and the performances do not depend on the qubit number $k$, which means that they will also be trivially parallelized to all qubits. The whole family of "bit flip " gates requires only one PS and one EOM.\\

 \section{Parallelization of qubit gate synthesis}
\label{result_para}

In this section, taking advantage of the possibility to address and manipulate a large number of modes with the same components E and P, we tackle the question of gate synthesis parallelization. More precisely, we explore the ability to parallelize the synthesis of an arbitrary single-qubit unitary, so that the same unitary can be applied to multiple qubits in parallel. We will moreover study the trade-offs between performance and the number of qubits on which the synthesis can be performed in parallel, both for frequency and time encodings and for the two three-component configurations, i.e. [EPE] and [PEP].

\begin{table}[]
\includegraphics[width=0.5\textwidth]{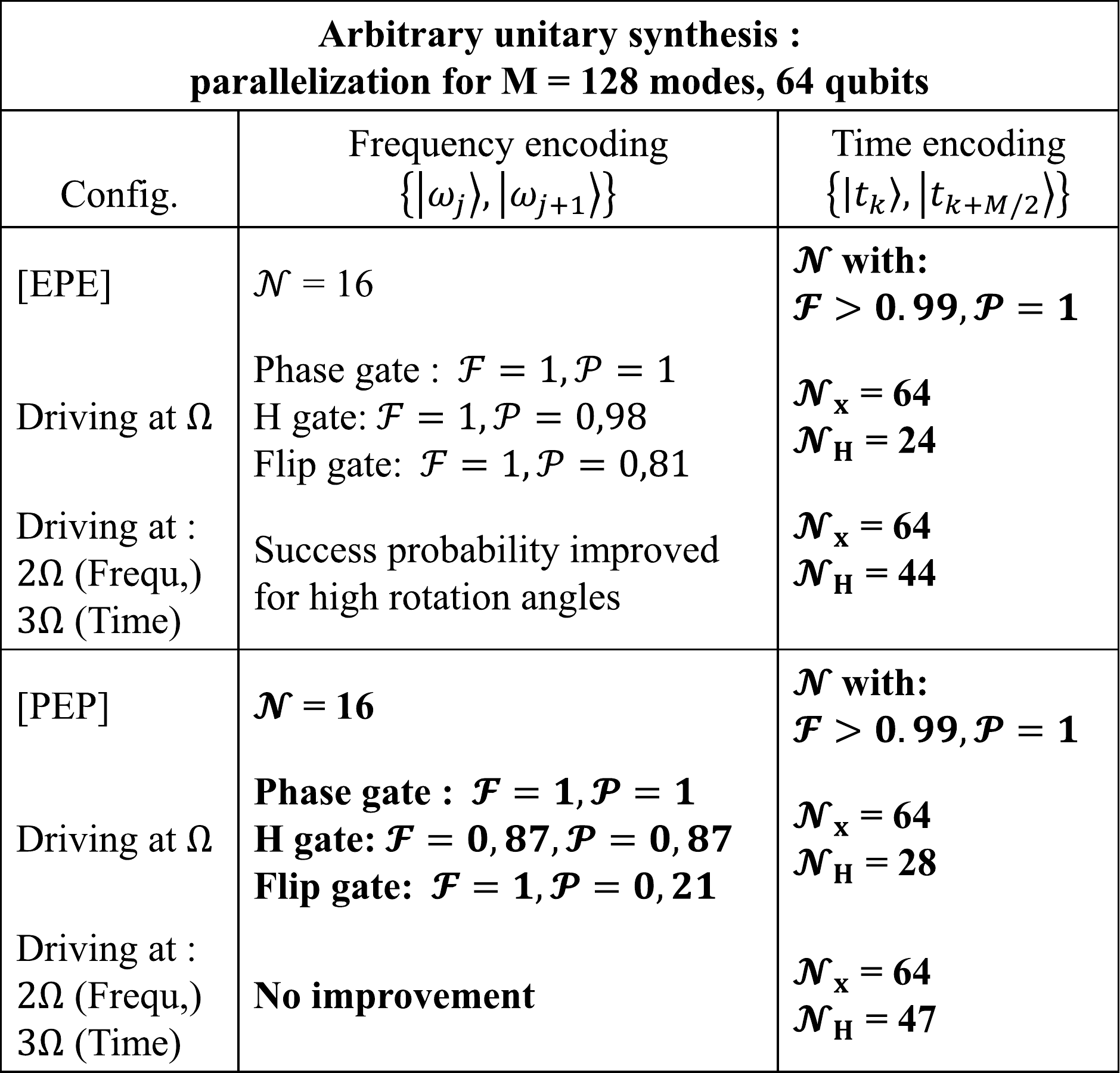}
\caption{Comparison of the possibilities of arbitrary unitary synthesis with [EPE] and [PEP] different configurations and for two kinds of encoding. $\mathcal{N}$ stands for the number of parallel qubits for which the gate fidelity and success probability are above threshold. The bold part corresponds to {\bf new results} reported in this work.}\label{AUS}
\end{table}
\normalsize
 \subsection{Frequency Encoding}
In the frequency basis, a qubit is encoded over two adjacent modes. Because the EOM inevitably scatters energy in the adjacent modes, using all the available frequency modes to encode quantum information would generate cross-talk problems.  To reduce such crosstalk, a spacing of 6 frequency modes  used as  buffer between two qubits has been reported in  \cite{lu2018electro}. It has lead to a success probability $>0.96$ in the [EPE] configuration, 

For the synthesis in the [PEP] configuration, that had not been investigated 
up to now, we compute the minimum mode spacing for which crosstalk is less than 0.001 between two nearest neighbour frequency qubits. We moreover chose to target the synthesis of the least favorable gate, namely the bit flip gate. Implementing the bit flip gate indeed requires a complete energy transfer between the two modes of the qubit, and hence a large modulation index. We obtained a minimum mode spacing of 6 frequency modes  as for the [EPE] configuration to reach $\mathcal{F}\geq 0.99\%$. Hence, a qubit effectively occupies 8 modes. The $0.1\%$ remaining energy leakage must be compared to the requirements of the considered transformations in order to evaluate the performance of this parallelization.


For a computational space composed of 128 frequency modes, which we consider for this analysis, a maximum number of $\frac{128}{8}=16$ qubit gates can thus be implemented in parallel for both [EPE] and [PEP] configurations with high performance (high fidelity and success probability).


\subsection{Time Encoding}
In section \ref{TE}, we showed that we were able to find the settings of the three components to synthesize with unit fidelity and success probability an arbitrary quantum gate for a single qubit encoded over two modes. The next step is to use all $128$ time modes to parallelize those transformations to a maximum number of qubits, the optimal number being $M/2=64$ qubits. 
\subsubsection{Multi-tone RF driving of the EOM}
\begin{center}
\fbox{\begin{minipage}{0.9\textwidth}
\begin{proposition}\label{EPE_temp_parallelization}
One can achieve an exact synthesis of an arbitrary unitary transformation U over all qubits with the [EPE] configuration with a square RF driving of the EOM.
\end{proposition}
\end{minipage}}
\end{center}
\begin{proof}
The parameters have to satisfy the following set of conditions
\begin{equation}
	\left\{ \begin{array}{cc}
        |\alpha| = \cos{(c/2)} \\
        |\beta| = \sin{(c/2)} 
    \end{array} \right. \\
    \forall k \in \llbracket 0, M/2-1 \rrbracket,
    \left\{ \begin{array}{cc}
        \phi_{1,k} =  \frac{ - d}{2} - \pi/4  \mod{\pi} \\
        \phi_{2,k} =  \frac{ - b}{2} + \pi/4  \mod{\pi} \\
        \phi_{1,k+M/2} = -\phi_{1,k}  \mod{\pi} \\
        \phi_{2,k+M/2} = -\phi_{2,k}  \mod{\pi}, \\
    \end{array} \right.
\end{equation}
which implies that the phases of the EOM are independent of time on the whole interval $[0, \frac{T}{2} ]=[ 0, \frac{\pi}{\Omega}]$:
The driving signal of the EOMs must then be of the form
\begin{equation}
x(t) = \mu\sum_{n=0}^{n=\infty} \frac{1}{2n+1} \sin{((2n+1)(\Omega t + \theta))},
\end{equation}
with $\mu = 1$ and $0<\theta<2\pi/M$.
\end{proof}
Although an infinite-sum square function allows arbitrary synthesis over all qubits, the practical implementation of such a signal is limited by the pass band of the EOMs and we consider therefore more realistic cases:

\begin{itemize}

\item The ideal square function can be truncated at order $N$
\begin{equation}
x(t) = \mu\sum_{n=0}^{N} \frac{1}{2n+1} \sin{((2n+1)(\Omega t + \theta))}.
\end{equation}
\item The truncated function can be optimized by using different modulation index $\mu$ and modulation phase $\theta$ for each frequency component
\begin{equation}
x(t) = \sum_{n=0}^{N} \frac{\mu_n}{2n+1} \sin{((2n+1)(\Omega t + \theta_n))}.
\end{equation}
\item The RF driving can be limited to a single frequency or very few frequency components, as will be considered in the next sections.
\end{itemize}


 
\begin{figure}[h!]
\centering\includegraphics[width=0.5\textwidth]{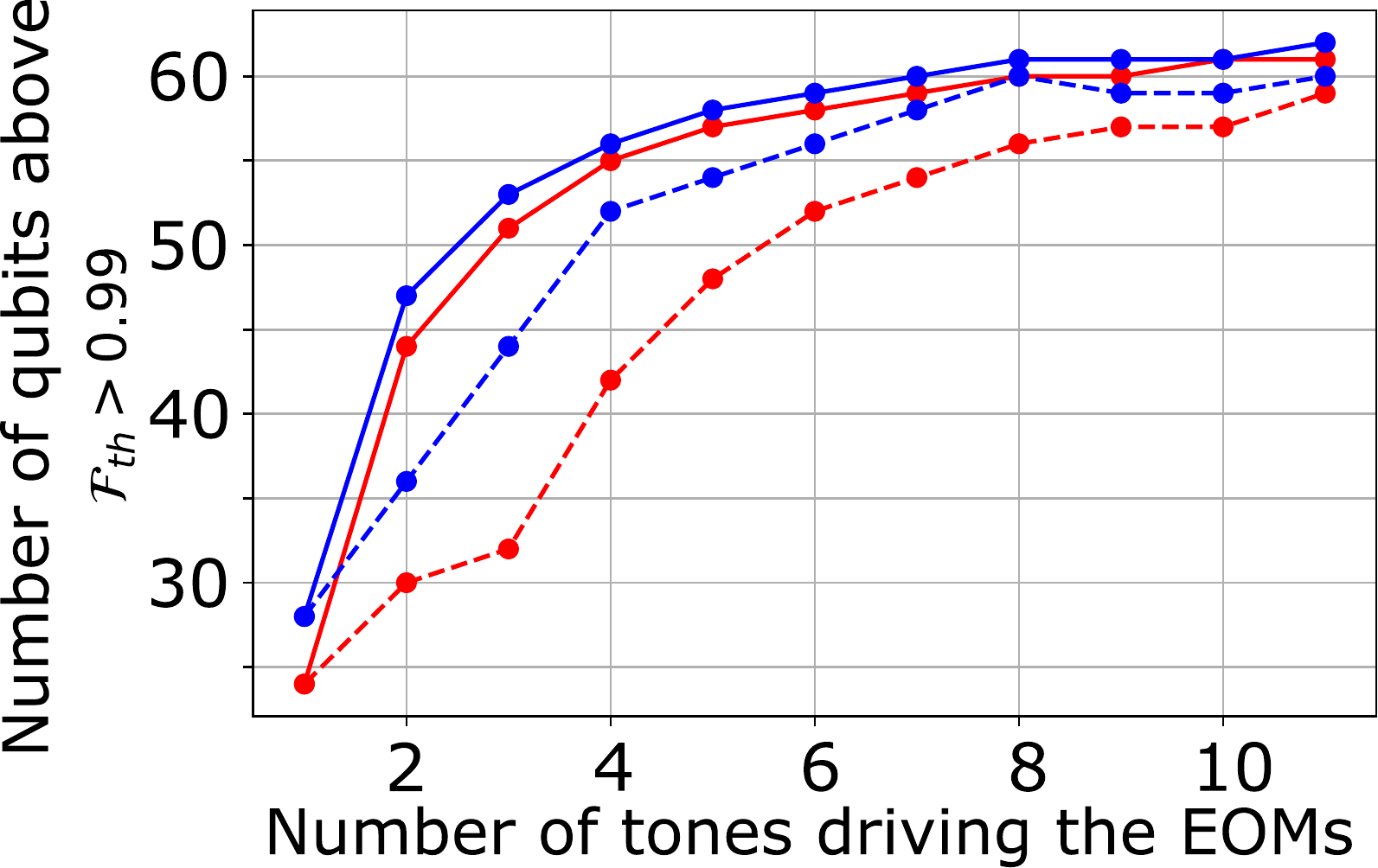}  
\caption{Parallel Hadamard gate synthesis {in time encoding framework}: number of qubits above fidelity threshold $\mathcal{F}_{th}=0.99$ as a function of the number of frequency components of the RF driving of the EOM, for the Hadamard transformation in [EPE] (red) and [PEP](blue) configurations with exact (dotted line) and optimized (full line) truncation of a square RF driving.}
\label{best_reachable_perf}
\end{figure}

For our numerical optimizations of the parallelization of gate synthesis, we use the fidelity and success probability acceptance thresholds\begin{equation}
         \mathcal{F}_{th}=0.99 \qquad
        \mathcal{P}_{th}=0.9999.
\end{equation}
Fig.\ref{best_reachable_perf} shows the number of parallel Hadamard gates above fidelity threshold as a function of the number of frequency tones in the RF driving signal of the EOMs for both [EPE] and [PEP] configurations. Considering 64 time-bin qubits, parallel synthesis of 28 Hadamard gates can be performed, (with $>99\%$ fidelity and almost $100\%$ success probability)  in the single-tone regime. This parallelization can be brought to  47 qubits with 2 tones, and 53 qubits with 3 tones. This positions spectral LOQC as a possible platform to conduct quantum metrology experiments where a large number of quantum modes must be manipulated to achieve better precision.\\

In the following sections, we derive the number of parallel transformations above a given performance threshold for the Hadamard and phase gates with a single tone RF driving of the EOM.
\subsubsection{Single-tone RF driving of the EOM}\label{single_tone_para}
\paragraph{Phase gates}

The matrix of phase gates is of the form
\begin{equation}
   \mathcal{M}=\left(\begin{array}{cc}
         1&0 \\
       0&e^{-i\nu}\\
    \end{array} \right)  .
\end{equation}

From the expression of the $W_k$ matrix corresponding to the [EPE] (red) and [PEP] (blue) configuration (see Eq. \ref{W_epe}), we derive a set of conditions on the component parameters for the parallelization, first for the [EPE] configuration
\begin{equation}
	\left\{ \begin{array}{c}
        |\alpha| = 1; \qquad
        |\beta| = 0 \\
       \forall k \in  [0, M/2-1]\qquad \phi_{1,k}+\phi_{2,k} =  \frac{\nu}{2} \mod{2\pi} \\
        \forall k \in  [0, M/2-1]\qquad
        \phi_{1,k+M/2}+ \phi_{2,k+M/2} = -\frac{\nu}{2} \mod{2\pi}\\
        \phi_{c_1}+\phi_{c_2}+\gamma=-\frac{\nu}{2} \mod{2\pi}.
    \end{array} \right. 
\end{equation}
As $\phi_{i,k+M/2}=-\phi_{i,k}$, we need only write
\begin{equation}
	\left\{ \begin{array}{c}
        |\alpha| = 1; \qquad
        |\beta| = 0 \\
       \forall k \in  [0, M/2-1]\qquad 
      \phi_{k,1}+\phi_{k,2}= \mu_1 \sin{(\frac{2k\pi}{M}+\theta_1)}
       +\mu_2 \sin{(\frac{2k\pi}{M}+\theta_2)} =  \frac{\nu}{2} \mod{2\pi} \\
        \phi_{c_1}+\phi_{c_2}+\gamma=-\frac{\nu}{2} \mod{2\pi}.
    \end{array} \right. 
\end{equation}
The two modulators act as one. It is then sufficient to consider only one modulator. We can write the previous set of equation as
\begin{equation}\label{epe_set}
	\left\{ \begin{array}{c}
        |\alpha| = 1; \qquad
        |\beta| = 0 \\
       \forall k \in  [0, M/2-1]\qquad 
      \phi_{k}= \mu \sin{(\frac{2k\pi}{M}+\theta)}
        =  \frac{\nu}{2} \mod{2\pi} \\
        \phi_{c}+\gamma=-\frac{\nu}{2} \mod{2\pi}.
    \end{array} \right. 
\end{equation}
In the same way, conditions can be found on the component parameters for the [PEP] configuration.
     \begin{equation}\label{pep1_set}
	\left\{ \begin{array}{c}
        |\alpha_1| = |\alpha_2| = 1; \qquad
        |\beta_1| =|\beta_2| = 0 \\
       \forall k \in  [0, M/2-1]\qquad 
       \phi_{k}=\mu \sin{(\frac{2k\pi}{M}+\theta)} =  \frac{\nu}{2} \mod{2\pi} \\
        \phi_{c}+\gamma_1+\gamma_2=-\frac{\nu}{2} \mod{2\pi}.
    \end{array} \right.
\end{equation}
or
\begin{equation}\label{pep2_set}
	\left\{ \begin{array}{c}
        |\alpha_1| = |\alpha_2| = 0; \qquad
        |\beta_1| =|\beta_2| = 1 \\
       \forall k \in  [0, M/2-1]\qquad 
       \phi_k = \mu \sin{(\frac{2k\pi}{M}+\theta)} = - \frac{\nu}{2} \mod{2\pi} \\
        \phi_{c}+\gamma_1+\gamma_2=-\frac{\nu}{2} \mod{2\pi}.
    \end{array} \right.
\end{equation}
These three sets of equations lead to the same synthesized matrix and therefore to the same formula for the fidelity. The fidelity of each qubit $k$ undergoing a phase change $\phi_k$ can be computed with the parameters for phase gates using Eq. \ref{eq:fidel} as
\begin{equation}
\label{fid_phase_k}
  \forall k \in [0, M/2-1]\qquad \mathcal{F}_k=\cos^2{\left(\phi_k-\frac{\nu}{2}\right)}.
\end{equation}

{Each qubit will not realize the gate with the same fidelity, as {this fidelity} depends on the value of the phase at a each time $k$. 
{We now compute the maximum number of qubits realizing the transformation with a fidelity better than a threshold $\mathcal{F}_{th}$
For each qubit $k$},
\begin{equation}
    \mathcal{F}_k\geq\mathcal{F}_{th} \Rightarrow \cos^2\left( \phi_k - \frac{\nu}{2} \right)\geq\mathcal{F}_{th}.
\end{equation}

Solving this inequation and replacing $\phi_k$ by its expression, we find

\begin{equation}
    \frac{\nu}{2}-\arccos\left( \sqrt{\mathcal{F}_{th}}\right) \leq \mu\sin\left(\frac{2\pi k}{M}+\theta\right) \leq \frac{\nu}{2}+\arccos\left( \sqrt{\mathcal{F}_{th}}\right).
    \label{eq:52}
\end{equation}

The maximum value that the sine can take is bounded by the upper bound. We thus set the modulation index to this upper bound $\mu = \frac{\nu}{2}+\arccos\left( \sqrt{\mathcal{F}_{th}}\right)$. Eq. (\ref{eq:52})  can then be rewritten as a condition on $k$:


\begin{equation}\label{final_phase_gate_fid}
    \frac{M}{2\pi}\arcsin \left( \frac{\frac{\nu}{2}-\arccos\left( \sqrt{\mathcal{F}_{th}}\right) }{\frac{\nu}{2}+\arccos\left( \sqrt{\mathcal{F}_{th}}\right) }  \right)-\theta \leq k\leq \frac{M}{2} - \frac{M}{2\pi}\arcsin \left( \frac{\frac{\nu}{2}-\arccos\left( \sqrt{\mathcal{F}_{th}}\right) }{\frac{\nu}{2}+\arccos\left( \sqrt{\mathcal{F}_{th}}\right) } \right)-\theta.
\end{equation}

The number $\mathcal{N}$ of qubit $k$ realizing the transformation with a fidelity over $\mathcal{F}_{th}$ is thus given by
\begin{equation}
    \mathcal{N} = \frac{M}{\pi}\left\lfloor \frac{\pi}{2}- \arcsin \left( \frac{\frac{\nu}{2}-\arccos\left( \sqrt{\mathcal{F}_{th}}\right) }{\frac{\nu}{2}+\arccos\left( \sqrt{\mathcal{F}_{th}}\right) } \right)\right\rfloor +1.
\end{equation}
}

\begin{figure}[h!]
\centering \includegraphics[width=0.5\textwidth]{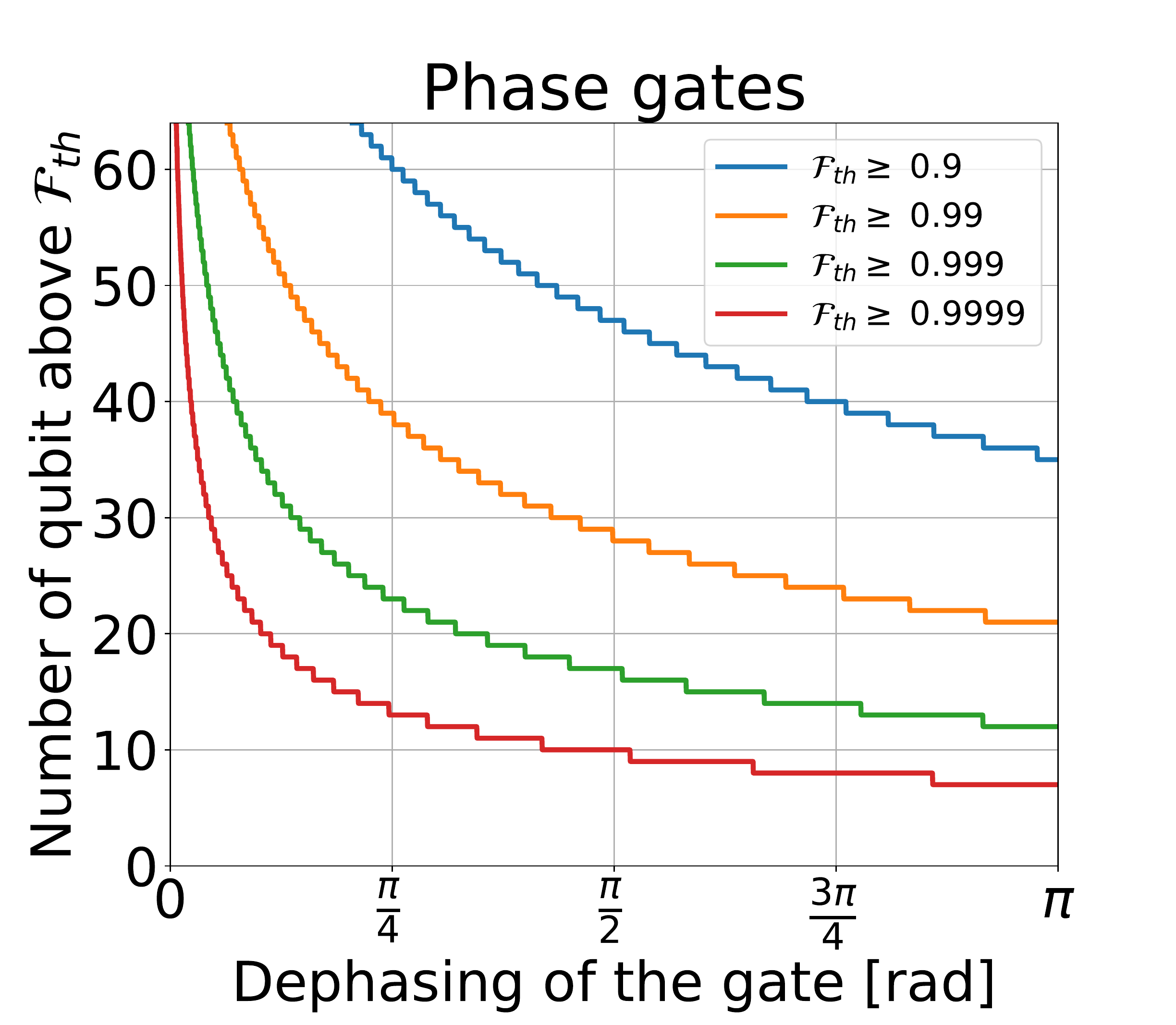}
    \caption{Phase gates parallel synthesis: maximum number of gates 
    {synthesized in parallel with a fidelity better than $\mathcal{F}_{th}$ }
    {by a phase gate operating on time modes (time encoding) as a function of the required gate z-rotation angle. This dependency holds for} both [EPE] and [PEP] configurations;
    Blue: $\mathcal{F}_{th}=0.9$; yellow: $\mathcal{F}_{th}=0.99$; green: $\mathcal{F}_{th}=0.999$; red: $\mathcal{F}_{th}=0.9999$.}
    \label{phase_perf_EPE_PEP}
\end{figure}

 $\mathcal{N}$ decreases with increasing $\nu$ showing that the limitation  comes from the phase of the EOM.
 The result is shown on Fig. \ref{phase_perf_EPE_PEP} for four values of the fidelity threshold.

\paragraph{Hadamard gates}
\label{Hadamard gates perf}
The Hadamard gate requiring both phase change and energy splitting between the two modes of the qubit subspace is therefore among the most difficult to synthesize.

Let us calculate the number of qubits realizing the Hadamard gate with performances exceeding $\mathcal{P}_{th}$ and $\mathcal{F}_{th}$ in [PEP] and [EPE] configurations:
the parameters corresponding to a single Hadamard gate for qubit $k$ are
\begin{equation}
	\left\{ \begin{array}{c}
        |\alpha| = |\beta| =\frac{1}{\sqrt{2}} \\
       \forall k \in [0, M/2-1]\qquad 
       \phi_{k} = \mu \sin{(\frac{2k\pi}{M}+\theta)}= \frac{\pi}{4} \mod{\pi}. \\
        
    \end{array} \right.
\end{equation}

By using the same method as for the phase gates, we can determine the fidelity achieved for each qubit $k$ as 
\begin{equation}\label{F_hadamard_epe}
    \forall k \in [0, M/2-1]\qquad \mathcal{F}_k^{(EPE)}=\sin^4\left(\phi_k+\frac{\pi}{4}\right)
\end{equation}
\begin{equation}\label{F_hadamard_pep}
    \forall k \in [0, M/2-1]\qquad \mathcal{F}_k^{(PEP)}=\sin^2\left(\phi_k+\frac{\pi}{4}\right)
\end{equation}
for respectively [EPE] and [PEP] configurations.

As for the phase gate, we look at the greater number of qubits that can realize the Hadamard transformation for [EPE] and [PEP] configurations. By starting with Eqs \ref{F_hadamard_epe} and \ref{F_hadamard_pep}, we find,

for the [EPE] configuration
\begin{equation}
    -\frac{\pi}{4}+\arcsin\left(\mathcal{F}_{th}^{1/4}\right) \leq \phi_k \leq 3\frac{\pi}{4}-\arcsin\left(\mathcal{F}_{th}^{1/4}\right),
\end{equation}
and for the [PEP] configuration
\begin{equation}
    -\frac{\pi}{4}+\arcsin\left(\mathcal{F}_{th}^{1/2}\right) \leq \phi_k \leq 3\frac{\pi}{4}-\arcsin\left(\mathcal{F}_{th}^{1/2}\right).
\end{equation}
{
Introducing  $\Delta\nu_{EPE} = \frac{\pi}{2} - \arcsin{({\mathcal{F}_{th}}^{1/4})}$    and $\Delta\nu_{PEP} = \frac{\pi}{2} - \arcsin{({\mathcal{F}_{th}}^{1/2})}$, this reads
\begin{equation}
    \frac{\pi}{4}-\Delta\nu_{EPE}\leq \phi_k \leq \frac{\pi}{4}+\Delta\nu_{EPE}, \qquad \frac{\pi}{4}-\Delta\nu_{PEP}\leq \phi_k \leq \frac{\pi}{4}+\Delta\nu_{PEP}. 
\end{equation}
Similarly to the phase gate calculation, we find the maximum number of qubits achieving the Hadamard transformation with a fidelity over $\mathcal{F}_{th}$ for both configurations} 
\begin{equation}\label{Nmax}
\mathcal{N_{EPE}} = \frac{M}{\pi}\left\lfloor
\frac{\pi}{2}-\arcsin{\Bigg(\frac{\frac{\pi}{4}-\Delta\nu_{EPE}}
{\frac{\pi}{4}+\Delta\nu_{EPE}}\Bigg)}\right\rfloor+1; \qquad
\mathcal{N_{PEP}} = \frac{M}{\pi}\left\lfloor
\frac{\pi}{2}-\arcsin{\Bigg(\frac{\frac{\pi}{4}-\Delta\nu_{PEP}}
{\frac{\pi}{4}+\Delta\nu_{PEP}}\Bigg)}\right\rfloor+1.
\end{equation}

\begin{center}
\fbox{\begin{minipage}{0.9\textwidth}
\begin{proposition}\label{EPEvsPEP}
For the Hadamard gate, the number of qubits realizing $\mathcal{F}_{th}$ is larger in the [PEP] configuration than in the [EPE] configuration for any finite number of frequencies in the RF driving.
\end{proposition}
\end{minipage}}
\end{center}
\begin{proof}
Let $k \in  [0, M/2-1]$ and $\mathcal{F}_{th}$ be a fidelity threshold.
\begin{equation}
    \mathcal{F}_k \geq \mathcal{F}_{th} \Leftrightarrow
    \phi_k \in {\big[\frac{\pi}{4}-\Delta\nu,\frac{\pi}{4}+\Delta\nu \big]}.
\end{equation}
In the [EPE]  configuration: $\Delta\nu_{EPE} = \frac{\pi}{2} - \arcsin{({\mathcal{F}_{th}}^{1/4})}$ \\
In the [PEP]  configuration: $\Delta\nu_{PEP} = \frac{\pi}{2} - \arcsin{({\mathcal{F}_{th}}^{1/2})}$ \\
$\Delta\nu_{PEP} \geq \Delta\nu_{EPE}$ \\
\end{proof}

\begin{figure}[h!]
\centering\includegraphics[width=0.5\textwidth]{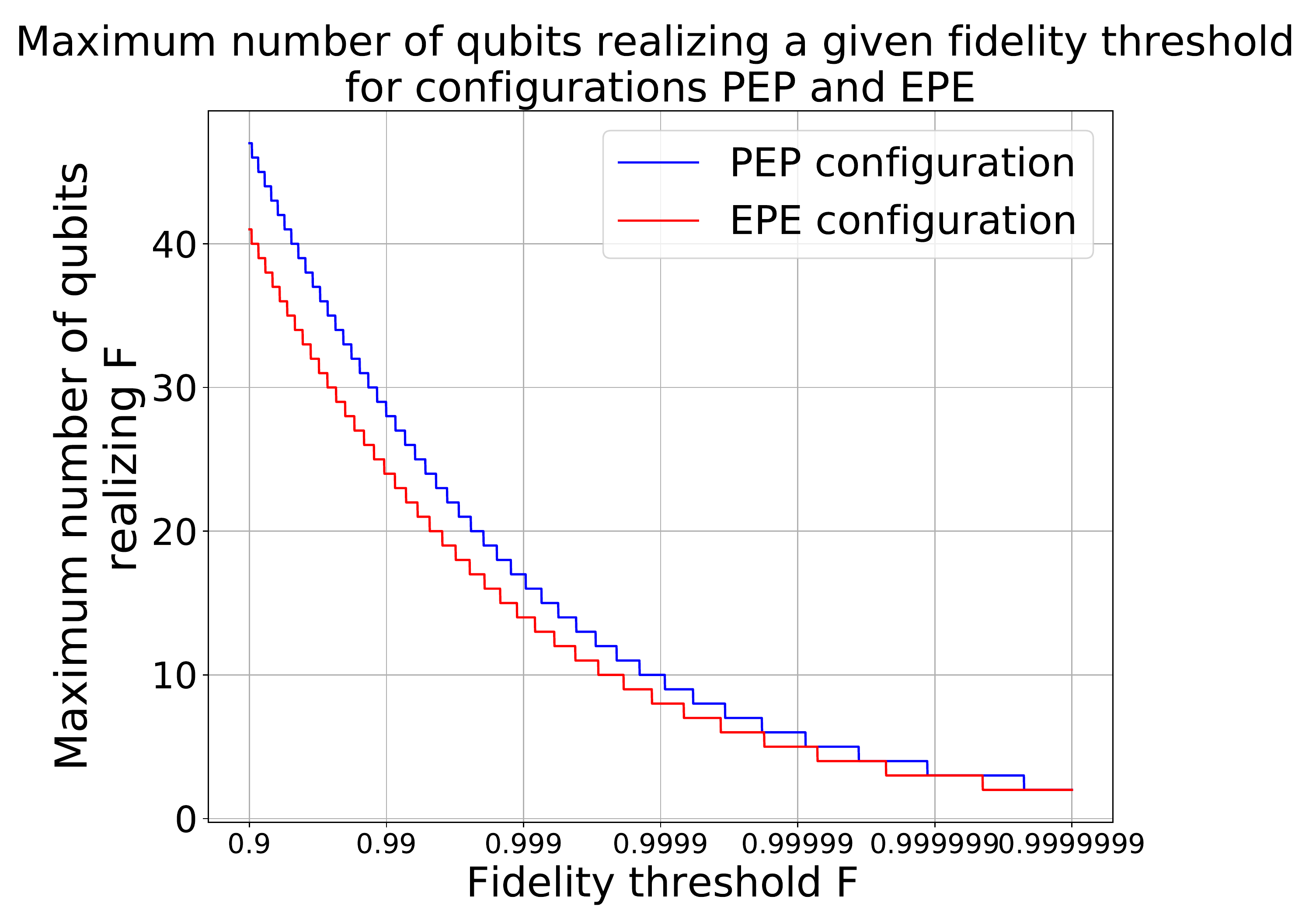}
\caption{Number of qubits with fidelity above threshold as a function of fidelity threshold for [EPE] (red) and [PEP] (blue) configurations with a single frequency RF modulation}
\label{nqb_of_F}
\end{figure}

Fig. \ref{nqb_of_F} {indeed} shows that the configuration [PEP] gives slightly better results than [EPE] for parallelization of the Hadamard gate.
It is noteworthy that the number of parallel phase gates corresponding to $\nu=\pi/2$ is 28, as in the case of the Hadamard gate suggesting that the performance limitation for the [PEP] configuration does not depend on the energy exchanges required by the gate but only on the required dephasing.

Table \ref{AUS} summarizes the main results on parallelization showing a clear advantage of time encoding over frequency encoding. This advantage of time encoding mainly comes from the two-scattering property (see Prop. \ref{2scat}) that can be used with the PS in the time basis, while it is not readily possible (under realistic RF bandwidth constraint) to use the EOM  n the frequency basis without coupling to frequency modes  outside of the Hilbert space, composed here of 2 neighboring frequencies, thereby imposing important constraints on the achievable success probability.
.


\color{black}

\section{Conclusions and Outlook}

 Optimized for the telecom industry, electro-optic modulators (E) and Pulse Shapers (P) components  perform mathematically complementary operations on multimode quantum photonic states.  Their combination gives rise to an appealing and engineering-friendly platform for photonic quantum information processing. P and E can in particular be used in combination to perform Spectral Linear Optics Quantum Computing (S-LOQC) in which quantum information encoding as well as measurements, are performed in the frequency basis \cite{lukens2017frequency}. Such S-LOQC platform has the ability to leverage high-dimensional frequency encodings and has proven to be a promising approach for photonic quantum information processing \cite{lu2018quantum, Lu_2020}.
 
In this work  we focus on a subset of S-LOQC tasks, namely the synthesis of an arbitrary single-qubit gate, and also the  parallelization of such synthesis. We moreover extend our investigation beyond standard S-LOQC by considering time-bin encoding, in addition to frequency encoding. 
We give  an overview of what can be achieved with the two paradigmatic minimal configurations,  [PEP] and [EPE],  for the processing of photonic qubits encoded either on frequency modes and on time modes, and we consider two main objectives: 
i)  the arbitrary unitary gate synthesis, with single tone driving of the EOMs
 ii)  the parallelization of the synthesis of a large number of unitaries with close-to-unit fidelity and success probability.\\
  Concerning i), we compute the achievable fidelities and probabilities of success for arbitrary unitaries with single-tone  RF driving of the EOMs. 
   Based on analytical and numerical studies, we show that time encoding allows the exact synthesis of an arbitrary single qubit unitary transformation with both configurations whereas frequency encoding only allows the exact synthesis of phase gates with maximal unit fidelity and success probability. \\
   Concerning ii), i.e. the parallelization,  we have compared the two configurations [PEP] and [EPE]  in terms of the number of qubit gates that can be synthesized in parallel with high performance, with a single setup of only 3 components with single- and dual-tone-driving of the EOMs. 
  In the time encoding framework, where exact synthesis is achievable, we have shown that the number of qubits that can be manipulated with high fidelity and probability is always higher than for frequency domain implementation. 
We also demonstrate that the [PEP] configuration performs slightly better than  [EPE]  to manipulate a larger number of time-mode-qubits.   We have moreover analyzed the impact of the addition of tones on the number of gates that can be synthesized in parallel and provide practical solutions -i.e. truncation strategies of the RF signal -  for high fidelity Hadamard and phase gates acting on multiple time modes.  
Quantitatively, considering a system with 128 addressable modes (and hence 64 qubits, with dual-rail encoding), that can be realistically addressed in the near-term with EOMs and PS, we notably exhibit, for time encoding, the possibility to perform the parallel synthesis of 28 Hadamard gates  (with $>99\%$ fidelity and almost $100\%$ success probability)  with a single-tone EOM driving. We also show that this parallelization can be extended to 47 qubits with 2 tones, and to 53 qubits with 3 tones.



Our work illustrates the interesting perspectives offered by the combined use of Pulse Shapers and RF-driven Electro-Optics Modulator to provide a versatile and powerful experimental platform for photonic information processing.
The immediate perspective opened by this work will consist in validating experimentally the theoretical results presented in the present article. Such experimental validation will require in particular to be able to perform measurements in the time basis, which constitutes a challenge, since the required time resolution (that would be around 10 ps  considering the orders of magnitude discussed in \ref{subsec:oom}) cannot be directly achieved using standard avalanche single photon detectors.  Time-resolved detection at the ps timescale will however be possible using superconducting nanowire \cite{korzh2020demonstration}, or by resorting to non-linear effects \cite{kuzucu2008time}. 
Another challenge will consist in designing experimental platforms operating over a large number of modes. We assumed a number of modes $M=128$ in the article, consistent with what is achievable with existing technology, and notably with pulse shaper frequency resolution. Leveraging  the possibility to integrate  electro-optic phase modulators and pulse shaper onto photonic chips \cite{nussbaum2022design,hu2021electro}  we can moreover expect to see a sharp increase in the number of addressable modes $M$ in the future, thereby boosting the information processing capabilities of such platform.

Performing quantum tomography typically requires the ability to perform arbitrary single qubit transformation followed by a measurement, and to parallelize such operation over a large number of qubits.  The [PEP] and [EPE] combinations considered in this article hence appear as an appealing experimental platform for quantum tomography over qubits, and could be extended to qudits,  with potential applications to the characterization of optical sources \cite{lu2021full}, or optical devices \cite{ansari2017temporal}. This platform can more generally be used to perform a high-dimensional Fourier Transform \cite{lu2022high, buddhiraju2021arbitrary}, which constitutes a central element in boson sampling or quantum metrology experiments. As seen in \ref{TE}, the [PEP] configuration used with time-encoded qubits indeed behaves as a linear interferometer that could for instance be used for quantum state engineering  \cite{tan2019resurgence} but also for optical interferometry with quantum-enhanced precision \cite{kaiser2018quantum}.
Our work highlights the relevance of considering time encoding to operate a quantum frequency processor composed of electro-optic modulators and pulse shapers. We have indeed shown that time-bin encoding enables the manipulation of a large number of qubits with high fidelity and high success probability and in general better performance that the frequency encoded versions. This work hence further positions the quantum frequency processor, that can readily leverage off-the-shelf device at low loss telecom wavelength,  as a promising contender for multimode quantum information processing.

\subsection*{Acknowledgements}

This work has been supported by Region Ile-de-France in the framework of DIM SIRTEQ, and has also benefited from the support of Telecom Paris Alumni.

\section*{APPENDICES}

\appendix
\section{ Generalised Expressions for Component Action in the Time Basis}\label{PS_temp}
Consider the action of a single PS in the time basis. For this, recall that in the spectral mode-basis, the PS-matrix has a representation of the form:
\begin{equation} \label{eq:1}
\Tilde{P} = 
\begin{bmatrix}
1 & 0 & \hdots & 0 \\
0 & e^{i\varphi_1} & \hdots & 0 \\
\vdots & \vdots & \ddots & \vdots \\ 
0 & 0 & \hdots & e^{i\varphi_{M-1}}
\end{bmatrix}
\end{equation}
where, $\varphi_1,\hdots,\varphi_{M-1}$ denote the real phases applied to the frequency modes.
The equivalent expression in the time basis:
\begin{multline}
   \Tilde{P}  = FPF^\dagger =
    \frac{1}{M}
    \begin{bmatrix}
    1 & 1 & 1 & \hdots & 1\\
    1 & \xi & \xi^2 & \hdots & \xi^{(M-1)}\\
    1 & \xi^2 & \xi^4 & \hdots & \xi^{2(M-1)}\\
    \vdots & \vdots & \vdots & \ddots & \vdots\\
    1 & \xi^{(M-1)} & \xi^{2(M-1)} & \hdots & \xi^{(M-1)^2}
    \end{bmatrix} \\
    \times\begin{bmatrix}
     1 & 0              &0& \hdots & 0 \\
     0 & e^{i\varphi_1} &0& \hdots & 0 \\
     0 &0               & e^{i\varphi_2} & \hdots & 0 \\
     \vdots & \vdots &   \vdots          & \ddots & \vdots \\ 
     0 & 0 &0& \hdots & e^{i\varphi_{M-1}}
    \end{bmatrix}
    \begin{bmatrix}
    1 & 1 & 1 & \hdots & 1\\
    1 & \xi^{-1} & \xi^{-2} & \hdots & \xi^{-(M-1)}\\
    1 & \xi^{-2} & \xi^{-4} & \hdots & \xi^{-2(M-1)}\\
    \vdots & \vdots & \vdots & \ddots & \vdots\\
    1 & \xi^{-(M-1)} & \xi^{-2(M-1)} & \hdots & \xi^{-(M-1)^2}
    \end{bmatrix}\notag
\end{multline}
\begin{multline}
     = \frac{1}{M}\begin{bmatrix}
    1 & e^{i\varphi_1} & e^{i\varphi_2} & \hdots & e^{i\varphi_{M-1}}\\
    1 & \xi e^{i\varphi_1} & \xi^2 e^{i\varphi_2} & \hdots & \xi^{(M-1)}e^{i\varphi_{M-1}}\\
    \vdots & \vdots & \vdots & \ddots & \vdots\\
    1 & \xi^{(M-1)}e^{i\varphi_1} & \xi^{2(M-1)} e^{i\varphi_2} & \hdots & \xi^{(M-1)^2} e^{i\varphi_{M-1}}
    \end{bmatrix}\\
    \times\begin{bmatrix}
    1 & 1 & 1 & \hdots & 1\\
    1 & \xi^{-1} & \xi^{-2} & \hdots & \xi^{-(M-1)}\\
    1 & \xi^{-2} & \xi^{-4} & \hdots & \xi^{-2(M-1)}\\
    \vdots & \vdots & \vdots & \ddots & \vdots\\
    1 & \xi^{-(M-1)} & \xi^{-2(M-1)} & \hdots & \xi^{-(M-1)^2}
    \end{bmatrix}
\end{multline}
where, $\xi = e^{i(2\pi/M)}$. This expression can be expressed more compactly as
\begin{equation}
   \forall k,k' \in  [0, M-1]\qquad
   P_{k'k} = \frac{1}{M}\sum_{j_1=0}^{M-1}\xi^{(k'-k)j_1}e^{i\varphi_{j_1}}.
\end{equation}

The matrix of the EOM is diagonal in the time basis
\begin{equation}
   \forall k',k'' \in [0, M-1]\qquad 
   E_{k''k'} = \delta_{k''k'}e^{i\phi_{k'}}
\end{equation}
The generic matrix element of the component combination [PEP] is therefore


\begin{gather}
    (\Tilde{P}E\Tilde{P})_{k'k} = \sum_j\bigg\{\frac{1}{M}\sum_{j_2}\xi^{(k'-j)j_2}e^{i\varphi_{j_2}}\bigg\}e^{i\phi_{j}}\bigg\{\frac{1}{M}\sum_{j_1}\xi^{(j-k)j_1}e^{i\varphi_{j_1}}\bigg\}\notag\\
    = \frac{1}{M^2}\sum_{j}e^{i\phi_{j}}\bigg[\sum_{j_2}\xi^{(k'-j)j_2}e^{i\varphi_{j_2}}\bigg]\bigg[\sum_{j_1}\xi^{(j-k)j_1}e^{i\varphi_{j_1}}\bigg]\notag\\
    = \frac{1}{M^2} \sum_{j,j_1,j_2} e^{i (\phi_{j}+\varphi_{j_1}+\varphi_{j_2}+\frac{2\pi}{M} [(j-k)j_1+(k'-j)j_2]}
  \end{gather}
  
  The corresponding expression for the [EPE] configuration is
 
  \begin{gather}
    (E\Tilde{P}E)_{k'k} =
    \bigg\{\sum_{k''}\delta_{k'''k''}e^{i\psi_{k''}}\bigg\}\bigg\{\frac{1}{M}\sum_{j}\omega^{(k''-k')j}e^{i\phi_{j}}\bigg\}\bigg\{\sum_{k}\delta_{k' k}e^{i\psi_{k}}\bigg\} \notag\\
       = \frac{1}{M}\sum_{j}e^{i\psi_{k"}}e^{i\frac{2\pi}{M}(k''-k')j}e^{i\phi_j}e^{i\psi_k}\notag\\
      =\frac{ e^{i(\phi_k+\phi_{k'})} }{M}\sum_{j} e^{i\frac{2\pi}{M}(k'-k)j} e^{i\varphi_j}
   \end{gather}

\section{Parameters of the PS as a two-scattering device in the time basis}\label{PS_temp_basis}

We start with no assumption on the "distance" $m$ between the two vectors of the qubit subspace 
\begin{equation}\label{2scatm}
   \forall k \in [0, M-m-1 ]\qquad \hat{U}_{PS}\ket{t_k}=\alpha\ket{t_k}+\beta\ket{t_{k+m}}
\end{equation}
where $0< m\leq M/2$ and $(\alpha, \beta) \in \mathbb{C}^{2}, |\alpha|^{2}+|\beta|^{2} = 1$
\\
The coupling constants between time modes can be derived from the Fourier transform of the pulse shaper action in the frequency domain
\begin{equation} 
    \forall (k,k') \in \{[ 0, M-1 ]\}^2, 
    P_{k',k} = \langle t_{k'}|\hat{U}_{PS}|t_k\rangle = \frac{1}{M}\sum_j \exp\left(i\frac{2\pi}{M}(k'-k)j+i\varphi_j\right)
\end{equation}
where $\varphi_j$ is the phase applied by the PS to the frequency mode $\ket{\omega_j}$ that we want to determine for $j\in [0, M-1]$.

Using Eq. \ref{2scatm} to compute $P_{k',k}$
\begin{equation}\label{eq_ps} 
	\left\{ \begin{array}{cc}
        \frac{1}{M} \sum_{j=0}^{M-1} e^{i\varphi_j} = \alpha \\
        \frac{1}{M} \sum_{j=0}^{M-1} e^{i 2 \pi \frac{mj}{M}}e^{i\varphi_j} = \beta \\
        \forall (p,q) \neq (k,k),(k+m,k),\qquad \frac{1}{M} \sum_{j=0}^{M-1} e^{i 2 \pi \frac{(p-q)j}{M}}e^{i\varphi_j} = 0
    \end{array} \right.
\end{equation}
Let $\,e^{i\varphi_j}=A_j\alpha+B_j\beta\, $.
Eq. (\ref{eq_ps}) becomes
\begin{equation}
\left\{
    \begin{array}{cc}
         \sum_{j=0}^{M-1} A_j\alpha = M\alpha \\
         \sum_{j=0}^{M-1} e^{i 2 \pi \frac{mj}{M}} B_j\beta = M\beta 
    \end{array}\right.
\end{equation}
These equations can be satisfied by choosing $A_j= 1$ and $ B_j= e^{-i 2 \pi \frac{mj}{M}} $,
which finally gives
\begin{equation} 
    \forall j \in [0, M-1],\qquad e^{i \varphi_j} = \alpha + \beta e^{-i 2 \pi \frac{mj}{M}}
\end{equation}

Then the unitarity of the PS matrix enforces the following relation
\begin{equation} 
    \forall j \in [0, M-1],\qquad |\alpha|^2 + |\beta|^2 + 2\Re  \mathfrak{e}\{ {\alpha \beta ^* e^{-i 2 \pi \frac{mj}{M}}} \} = 1
\end{equation}
The case $j=0$ imposes $\alpha \beta ^*$ to be purely imaginary. Therefore we can set $\alpha = |\alpha|e^{i\gamma}$ and $\beta = \pm i|\beta|e^{i\gamma}$, which yields a new unitarity condition
\begin{equation} 
    \forall j \in [0, M-1], \qquad \Im \mathfrak{m} \{ { e^{-i 2 \pi \frac{mj}{M}}} \} = 0
\end{equation}
which sets $m = M/2$. \\
It is therefore possible to choose settings of the PS so that it scatters the energy of mode $\ket{t_k}$ only into the two modes $\ket{t_k}$ and $\ket{t_{k+M/2}}$. In the frequency basis
\begin{equation}
  \forall j \in [0, M-1], \qquad \hat{U}_{PS}\ket{\omega_j}=\left[\alpha+\beta e^{-i \pi j}\right]\ket{\omega_j}
\end{equation}
Finally, we obtain
\begin{equation}
\left\{
\begin{array}{cc}
    \forall k\in [0, M/2], \hat{U}_{PS}\ket{t_k}=\alpha\ket{t_k}+\beta\ket{t_{k+M/2}}\\
    \forall k\in [M/2, M-1], \hat{U}_{PS}\ket{t_k}=\beta\ket{t_k}+\alpha\ket{t_{k-M/2}}.
\end{array}
\right.
\end{equation}
Subsequently the PS matrix takes the following form in the time basis:
\begin{equation}
    \tilde{PS} =  e^{i\gamma}
    \left( \begin{array}{ccc}
    |\alpha|I & \pm i|\beta|I \\
    \pm i|\beta|I & |\alpha|I 
    \end{array} \right)
\end{equation}
where $I$ is the identity matrix of dimension $M/2$.
\color{black}

\bibliography{bibmultimode20Oct}
\end{document}